\documentclass[twocolumn,showpacs,preprintnumbers,amsmath,amssymb,showkeys]{revtex4}

\usepackage{graphicx}
\usepackage{dcolumn}
\usepackage{bm}

\begin{document}

\preprint{APS/123-QED}

\newcounter{univ_counter}
\setcounter{univ_counter} {0}
\addtocounter{univ_counter} {1} 
\edef\MSU{$^{\arabic{univ_counter}}$ } \addtocounter{univ_counter} {1} 
\edef\INFNGE{$^{\arabic{univ_counter}}$ } \addtocounter{univ_counter} {1} 
\edef\ROMATRE{$^{\arabic{univ_counter}}$ } \addtocounter{univ_counter} {1} 
\edef\RPI{$^{\arabic{univ_counter}}$ } \addtocounter{univ_counter} {1} 
\edef\SACLAY{$^{\arabic{univ_counter}}$ } \addtocounter{univ_counter} {1} 
\edef\WM{$^{\arabic{univ_counter}}$ } \addtocounter{univ_counter} {1} 
\edef\UMASS{$^{\arabic{univ_counter}}$ } \addtocounter{univ_counter} {1} 
\edef\JLAB{$^{\arabic{univ_counter}}$ } \addtocounter{univ_counter} {1} 
\edef\INFNFR{$^{\arabic{univ_counter}}$ } \addtocounter{univ_counter} {1} 
\edef\YEREVAN{$^{\arabic{univ_counter}}$ } \addtocounter{univ_counter} {1} 
\edef\ASU{$^{\arabic{univ_counter}}$ } \addtocounter{univ_counter} {1}
\edef\ORSAY{$^{\arabic{univ_counter}}$ } \addtocounter{univ_counter} {1} 
\edef\EDINBURGH{$^{\arabic{univ_counter}}$ } \addtocounter{univ_counter} {1} 
\edef\GWU{$^{\arabic{univ_counter}}$ } \addtocounter{univ_counter} {1} 
\edef\UNH{$^{\arabic{univ_counter}}$ } \addtocounter{univ_counter} {1} 
\edef\OHIOU{$^{\arabic{univ_counter}}$ } \addtocounter{univ_counter} {1} 
\edef\CMU{$^{\arabic{univ_counter}}$ } \addtocounter{univ_counter} {1} 
\edef\CUA{$^{\arabic{univ_counter}}$ } \addtocounter{univ_counter} {1} 
\edef\SCAROLINA{$^{\arabic{univ_counter}}$ } \addtocounter{univ_counter} {1} 
\edef\ODU{$^{\arabic{univ_counter}}$ } \addtocounter{univ_counter} {1} 
\edef\UTEP{$^{\arabic{univ_counter}}$ } \addtocounter{univ_counter} {1} 
\edef\VIRGINIA{$^{\arabic{univ_counter}}$ } \addtocounter{univ_counter} {1} 
\edef\PITT{$^{\arabic{univ_counter}}$ } \addtocounter{univ_counter} {1} 
\edef\FSU{$^{\arabic{univ_counter}}$ } \addtocounter{univ_counter} {1} 
\edef\CNU{$^{\arabic{univ_counter}}$ } \addtocounter{univ_counter} {1} 
\edef\VT{$^{\arabic{univ_counter}}$ } \addtocounter{univ_counter} {1} 
\edef\DUKE{$^{\arabic{univ_counter}}$ } \addtocounter{univ_counter} {1} 
\edef\UCONN{$^{\arabic{univ_counter}}$ } \addtocounter{univ_counter} {1} 
\edef\MIT{$^{\arabic{univ_counter}}$ } \addtocounter{univ_counter} {1} 
\edef\URICH{$^{\arabic{univ_counter}}$ } \addtocounter{univ_counter} {1} 
\edef\JMU{$^{\arabic{univ_counter}}$ } \addtocounter{univ_counter} {1} 
\edef\GLASGOW{$^{\arabic{univ_counter}}$ } \addtocounter{univ_counter} {1} 
\edef\NSU{$^{\arabic{univ_counter}}$ } \addtocounter{univ_counter} {1} 
\edef\KYUNGPOOK{$^{\arabic{univ_counter}}$ } \addtocounter{univ_counter} {1} 
\edef\ITEP{$^{\arabic{univ_counter}}$ } \addtocounter{univ_counter} {1} 
\edef\RICE{$^{\arabic{univ_counter}}$ } \addtocounter{univ_counter} {1} 
\edef\FIU{$^{\arabic{univ_counter}}$ } \addtocounter{univ_counter} {1} 
\edef\UCLA{$^{\arabic{univ_counter}}$ } \addtocounter{univ_counter} {1} 

\title{
A Kinematically Complete Measurement of the Proton Structure Function $\mathbf F_2$
in the Resonance Region and Evaluation of Its Moments\\}

\author{
M.~Osipenko,\MSU$^,$\INFNGE\
G.~Ricco,\INFNGE\
M.~Taiuti,\INFNGE\
M.~Ripani,\INFNGE\
S.~Simula,\ROMATRE\ 
G.~Adams,\RPI\
E.~Anciant,\SACLAY\
M.~Anghinolfi,\INFNGE\
B.~Asavapibhop,\UMASS\
G.~Audit,\SACLAY
% ...
\
\\(The CLAS Collaboration)}

\address{\MSU Moscow State University, 119992 Moscow, Russia}
\address{\INFNGE INFN, Sezione di Genova, and Dipartimento di Fisica dell'Universit\`a, 16146 Genova, Italy}
\address{\ROMATRE Universit\`a di ROMA III, 00146 Roma, Italy}
\address{\RPI Rensselaer Polytechnic Institute, Troy, New York 12180}
\address{\SACLAY CEA-Saclay, Service de Physique Nucl\'eaire, F91191 Gif-sur-Yvette, Cedex, France}
\address{\WM College of William and Mary, Williamsburg, Virginia 23187}
\address{\UMASS University of Massachusetts, Amherst, Massachusetts 01003}
\address{\JLAB Thomas Jefferson National Accelerator Facility, Newport News, Virginia 23606}
\address{\INFNFR INFN, Laboratori Nazionali di Frascati, PO 13, 00044 Frascati, Italy}
\address{\YEREVAN Yerevan Physics Institute, 375036 Yerevan, Armenia}
\address{\ASU Arizona State University, Tempe, Arizona 85287}
\address{\ORSAY Institut de Physique Nucleaire ORSAY, IN2P3 BP 1, 91406 Orsay, France}
\address{\EDINBURGH Edinburgh University, Edinburgh EH9 3JZ, United Kingdom}
\address{\GWU The George Washington University, Washington, DC 20052}
\address{\UNH University of New Hampshire, Durham, New Hampshire 03824}
\address{\OHIOU Ohio University, Athens, Ohio 45701}
\address{\CMU Carnegie Mellon University, Pittsburgh, Pennsylvania 15213}
\address{\CUA Catholic University of America, Washington, D.C. 20064}
\address{\SCAROLINA University of South Carolina, Columbia, South Carolina 29208}
\address{\ODU Old Dominion University, Norfolk, Virginia 23529}
\address{\UTEP University of Texas at El Paso, El Paso, Texas 79968}
\address{\VIRGINIA University of Virginia, Charlottesville, Virginia 22901}
\address{\PITT University of Pittsburgh, Pittsburgh, Pennsylvania 15260}
\address{\FSU Florida State University, Tallahassee, Florida 32306}
\address{\CNU Christopher Newport University, Newport News, Virginia 23606}
\address{\VT Virginia Polytechnic Institute and State University, Blacksburg, Virginia   24061}
\address{\DUKE Duke University, Durham, North Carolina 27708}
\address{\UCONN University of Connecticut, Storrs, Connecticut 06269}
\address{\MIT Massachusetts Institute of Technology, Cambridge, Massachusetts  02139}
\address{\URICH University of Richmond, Richmond, Virginia 23173}
\address{\JMU James Madison University, Harrisonburg, Virginia 22807}
\address{\GLASGOW University of Glasgow, Glasgow G12 8QQ, United Kingdom}
\address{\NSU Norfolk State University, Norfolk, Virginia 23504}
\address{\KYUNGPOOK Kyungpook National University, Taegu 702-701, South Korea}
\address{\ITEP Institute of Theoretical and Experimental Physics, Moscow, 117259, Russia}
\address{\RICE Rice University, Houston, Texas 77005}
\address{\FIU Florida International University, Miami, Florida 33199}
\address{\UCLA University of California at Los Angeles, Los Angeles, California  90095}


\date{\today}

\begin{abstract}
We measured the inclusive electron-proton cross section in the nucleon
resonance region ($W<2.5$~GeV) at momentum transfers $Q^2$ below
$4.5$~(GeV/c)$^2$ with the CLAS detector. The 
large acceptance of CLAS allowed
the measurement of the cross section in
a large, contiguous two-dimensional range of $Q^2$ and $x$, making it possible to perform
an integration of the data at fixed $Q^2$ over the significant
$x$-interval. From these
data we extracted the structure function $F_2$ and, by including other
world data, we studied the $Q^2$ evolution of its moments, $M_n(Q^2)$, in order
to estimate higher twist contributions.
The small statistical and systematic uncertainties of the CLAS data allow a
precise extraction of the higher twists and will require significant improvements
in theoretical predictions if a meaningful comparison with these new
experimental results is to be made.
\end{abstract}

\pacs{12.38.Cy, 12.38.Lg, 12.38.Qk, 13.60.Hb}

\keywords{moments, nucleon structure, higher twists, QCD, OPE}

\maketitle

\section{\label{sec:introduction}Introduction}
The striking features of the nucleon structure function $F_2$
were first noted nearly 30 years ago by Bloom and
Gilman~\cite{BloGil}. They empirically observed two effects in data measured
at SLAC: a) the dual behaviour of the $F_2(x,Q^2)$
function that shows common features between the two kinematic regions
corresponding to the nucleon resonances and Deep Inelastic Scattering
(DIS), b) the extension of the scaling region to lower $Q^2$ values
when $F_2(x',Q^2)$ is plotted as a function of $x'=x/(1+M^2x^2/Q^2)$,
the ``improved scaling variable''. More precisely, they found
that the smooth function $F_2(x')$ measured at high $Q^2$ in the DIS
region represents a good average over the resonances of the
$F_2(x',Q^2)$ structure function measured at lower $Q^2$ values.
Moreover, the duality appears to be valid locally. In fact, each of
the most prominent resonance bumps, when averaged within its width, shows
approximate scaling~\cite{f2mom-hc}. Later on, in the framework of QCD,
De Rujula, Georgi and Politzer~\cite{Rujula} provided the first explanation
of the Bloom-Gilman duality. They evaluated the Cornwall-Norton~\cite{Corn-Nort}
moments $M_n^{CN}(Q^2)$ of the nucleon structure function $F_2$, defined as
\begin{equation}
M^{CN}_n(Q^2)=\int_0^1dxx^{(n-2)}F_2(x,Q^2),
\label{eq:CWMoment}
\end{equation}
\noindent
using the Operator Product Expansion (OPE) they obtained the
following expression
\begin{equation}
M^{CN}_n(Q^2)=A_n(Q^2)+\sum_{k=1}^{\infty}\biggl(n\frac{\gamma^2}
{Q^2}\biggr)^kB_{nk}(Q^2),
\label{eq:CWMomentExp}
\end{equation}
\noindent
where $A_n(Q^2)$ can be evaluated in the framework of perturbative
QCD (pQCD), and it is directly connected to the corresponding moment
of the asymptotic limit of $F_2$. The contribution of the higher twists,
which is related to multi-parton correlations
inside the nucleon
and represented by $B_{nk}(Q^2)$,
depends on the value of the constant $\gamma^2$ in such a way that
$\gamma^2$ can be considered as
a scale constant for higher twist effects. Assuming a small
value of the constant $\gamma^2$, the authors of Ref.~\cite{Rujula}
showed that the contribution of the higher twists was relatively small,
at least for low values of $n$ and for $Q^2 \ge M^2$, justifying
the observed dual behaviour of the structure function.
 
It is now well established that the interpretation of the parton-hadron
duality in light of QCD requires the evaluation of the moments of
the nucleon structure functions and their evolution as a function of
$Q^2$. Current pQCD calculations can estimate the $Q^2$ evolution up
to the Next-to-Next-to-Leading Order, giving access to the
interesting kinematic region of high $x$ and moderate $Q^2$ where
the multi-parton correlation contribution to the nucleon wave function
becomes dominant.

The interest in investigating multi-parton correlations
in inelastic lepton scattering off the nucleon at large
values of $x$ has recently been renewed, leading to a re-analysis of old $F_2$
data~\cite{Ricco2,Nichtw}. Unfortunately, the results from
Ref.~\cite{Ricco2} and \cite{Nichtw} were mainly based on the
analysis of fits of the structure function $F_2$ and therefore
were still qualitative.
Moreover, the previous lack of data in the resonance region did not allow
a model independent evaluation of the moments' evolution to lower $Q^2$,
and therefore offered very few opportunities to
quantitatively investigate the role of QCD below the DIS limit.

The Hall~C Collaboration at the Thomas Jefferson National Accelerator Facility (TJNAF) has
recently provided high quality data in this kinematic region~\cite{f2-hc},
allowing a more precise evaluation of the moments of the $F_2$ structure
function of the proton~\cite{f2mom-hc}.
However, like many other such measurements, the data were taken with a
spectrometer of relatively small angular acceptance and the measured
inclusive cross sections do not span a large continuous $x$ interval for
constant $Q^2$.  Data taken in this manner follow a kinematic locus in $Q^2$
vs. $x$ and require substantial interpolation to determine the $F_2$ moments.
\begin{figure}
\includegraphics[bb=1cm 6cm 20cm 23cm, scale=0.4]{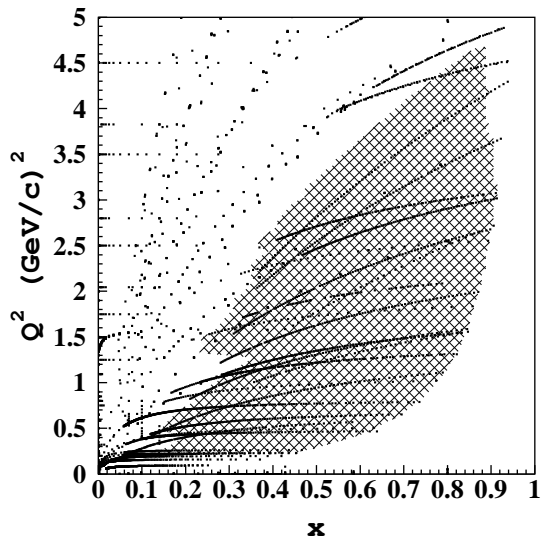}
\caption{\label{fig:xandQ2Domain} Experimental data on the structure
function $F_2(x,Q^2)$ used for the moment evaluation in the CLAS
kinematic region: points - world data; shaded area - CLAS data.}
\end{figure}

In this paper we report the first measurement in a wide continuous
interval in $x$ and $Q^2$ (see Fig.~\ref{fig:xandQ2Domain})
of the inclusive electron-proton scattering
cross section. These measurements were performed at TJNAF with the CLAS detector in Hall B.
The $F_2$ structure function was extracted over the whole resonance
region ($W \le 2.5$~GeV) below $Q^2 = 4.5$~(GeV/c)$^2$. This measurement,
together with existing world data, allowed for the evaluation of
the $F_2$ moments, drastically reducing the uncertainties related to
data interpolation and providing the most detailed dependence on $Q^2$
of the moments up to $n=8$. Furthermore, the elastic contribution to the
moments was updated with respect to Ref.~\cite{Ricco2} using the fit
of the nucleon form factors from Ref.~\cite{Bosted} adjusted to the Jefferson Lab data on
the ratio $G_E/G_M$~\cite{hall-a}, as described in Ref.~\cite{CLAS_note}.
Finally, we used our new determination of the $F_2$ moments to extract
the higher twist contribution as a function of $Q^2$.

In section~\ref{sec:Moments} we review
the $F_2$ moments in the framework of pQCD.
In section~\ref{sec:DataAnalysis} we discuss the analysis of the data,
including the extraction of the $F_2$ structure function from the
cross section. The evaluation of the moments and uncertainties is
also presented in section~\ref{sec:DataAnalysis}. Finally,
Section~\ref{sec:Discussion} is devoted to the interpretation of the results.

\section{Moments of the Structure Function $F_2$}\label{sec:Moments}
Until recently the studies of inclusive lepton-nucleon scattering
represented the main source of information about nucleon structure.
In the DIS region, measured structure functions can be directly connected
to the parton momentum distribution of the nucleon in the framework of pQCD.
After the successful interpretation of the DIS region, the intermediate
kinematic domain, situated at $Q^2$ of a few (GeV/c)$^2$ and large values
of $x$, attracted the interest of physicists.
Despite interpretation difficulties, this region allows the study of
multi-parton correlation contribution to the proton wave function.
These processes are not accessible in
DIS due to the small value of the running coupling constant $\alpha_S(Q^2)$.

The OPE of the virtual photon-nucleon scattering
amplitude leads to the description of the complete $Q^2$ evolution of the
moments of the nucleon structure functions. The n-th Cornwall-Norton moment~\cite{Corn-Nort}
of the (asymptotic) structure function $F_2(x,Q^2)$ for a massless nucleon
can be expanded as:
\begin{equation}
M^{CN}_n(Q^2)=\sum_{\tau=2k}^{\infty}E_{n \tau}(\mu,Q^2)
O_{n \tau}(\mu)\biggl(\frac{\mu^2}{Q^2}\biggr)^{{1 \over 2}(\tau-2)},
\label{eq:i_m1}
\end{equation}
\noindent
where $k=1,2,...,\infty$, $\mu$ is the factorization scale,
$O_{n \tau}(\mu)$ is the reduced matrix element of the local operators
with definite spin $n$ and twist $\tau$ (dimension minus spin), related to
the non-perturbative structure of the target.  $E_{n \tau}(\mu,Q^2)$
is a dimensionless coefficient function describing the small distance behaviour,
which can be perturbatively expressed as a power expansion of the running
coupling constant $\alpha_s(Q^2)$.

At $Q^2$ values comparable with the squared proton mass, $M^2$, the structure
function $F_2$ still contains non-negligible mass-dependent terms that
produce in Eq.~\ref{eq:i_m1} additional $M^2/Q^2$ power corrections
(kinematic twists). To avoid these terms, the moments $M^{CN}_n(Q^2)$ of
the massless $F_2$ have to be replaced in Eq.~\ref{eq:i_m1}
by the corresponding Nachtmann~\cite{Nachtmann} moments $M^{N}_n(Q^2)$ of
the measured structure function $F_2(x,Q^2)$ (see also Ref.~\cite{Roberts}),
It has been shown that:
\begin{equation}
M^{CN}_n\bigl(F^{lim}_2(x,Q^2)\bigr)=M^{N}_n\bigl(F_2(x,Q^2)\bigr),
\end{equation}
\noindent
where $F^{lim}_2(x,Q^2)$ is the asymptotic structure function of
the massless nucleon and
\begin{eqnarray}\nonumber
M^{N}_n(Q^2) = &&\int_0^1 dx \frac{\xi^{n+1}}{x^3} F_2(x,Q^2)\\ 
&&\Biggl[\frac{3+3(n+1)r+n(n+2)r^2}{(n+2)(n+3)}\Biggr],
\label{eq:i_nm1}
\end{eqnarray}
\noindent
where $r = \sqrt{1+4M^2x^2/Q^2}$ and $\xi = 2x/(1+r)$.

Since the moments in Eq.~\ref{eq:i_m1} are totally inclusive, the elastic
contribution at $x=1$ has to be added according to Ref.~\cite{f2mom-hc}:
\begin{equation}
F^{el}_2(x,Q^2)= \delta (1-x) \frac{\bigl(G_E^2(Q^2)+
\frac{Q^2}{4M^2}G_M^2(Q^2)\bigr)}{\bigl(1+\frac{Q^2}{4M^2}\bigr)},
\end{equation}
\noindent
with $G_E^2$ $(G_M^2)$ being the proton electric (magnetic) elastic form factor.

\begin{figure}
\includegraphics[bb=1cm 12cm 20cm 19cm, scale=0.5]{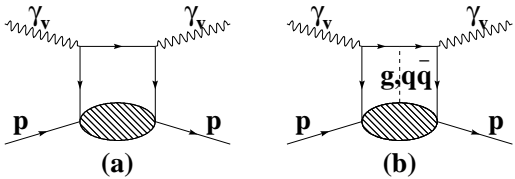}
\caption{\label{fig:TwistDiag} Twist diagrams:(a) the leading
twist contribution evaluated at leading order of pQCD; (b) the
contribution of higher twists, where current quark and nucleon remnant can
exchange by a system of particles consisting of gluons and $q\bar{q}$-pairs
whose complexity is increasing with twist order.}
\end{figure}

For the leading $\tau=2$ twist, one ends up in Leading Order (LO) or 
Next-to-Leading Order (NLO)
with the well known perturbative logarithmic $Q^2$ evolution of singlet
and non-singlet $F_2$ moments. However, if one wants to extend the
analysis to small $Q^2$ and large $x$ where the rest of the perturbative
series becomes significant, some procedure for the summation of the higher
orders of the pQCD expansion, such as the infrared renormalon model~\cite{renormalon,Ricco1} or
the recently developed soft-gluon resummation technique~\cite{SGR,SIM00}, has to be applied.
For higher twists, $\tau >2$, the power terms $E_{n \tau}(\mu)$ are
related to quark-quark and quark-gluon correlations, as illustrated
by Fig.~\ref{fig:TwistDiag}, and should become important
at small $Q^2$.

The systematic analysis of the $Q^2$ dependence of the experimentally derived
Nachtmann moments $M^{N}_n(Q^2)$ in the intermediate $Q^2$ range
($0.5<Q^2<10$~(GeV/c)$^2$) should allow a separation of the
higher twists from the leading twist. A precise evaluation would permit
a comparison with the QCD predictions obtained from lattice simulations or
a comparison with those models that describe the non-perturbative domain. 

\section{Data Analysis}\label{sec:DataAnalysis}
The data were collected at TJNAF in Hall B with the CLAS detector
and a liquid hydrogen target with thickness $\rho x=0.35$~g/cm$^2$
during the electron beam running period in
February-March 1999.
The average beam current of 4.5 nA corresponded to a luminosity
of 6x10$^{33}$ cm$^{-2}$s$^{-1}$.
To cover the largest interval in $Q^2$ and $x$,
data were taken at five different electron beam energies:
$E_0=$~1.5, 2.5, 4.0, 4.2 and 4.4 GeV.
The CLAS detector is a magnetic spectrometer based on a six-coil
torus magnet whose field is primarily oriented along the azimuthal
direction. The sectors, located between the magnet coils, are instrumented
individually to form six essentially independent magnetic spectrometers. The particle
detection system includes drift chambers (DC) for track
reconstruction~\cite{dc},
scintillation counters (TOF) for the time of flight measurement~\cite{sc},
Cherenkov counters (CC) for electron identification~\cite{cc}, and
electromagnetic calorimeters (EC) to measure neutrals and to improve the
electron-pion separation~\cite{ec}.  The EC detectors have
a granularity defined by triangular cells in the plane perpendicular
to the incoming particles to study the electromagnetic shower shape and
are longitudinally divided into two parts with the inner part acting as
a pre-shower. Charged particles can be detected and identified for momenta
down to 0.2 GeV/c and for polar angles between 8$^\circ$ and 142$^\circ$.
The CLAS superconducting coils reduce the acceptance of about 80\%
at $\theta=90^{\circ}$ to about 50\% at forward angles ($\theta=20^{\circ}$),
while the total acceptance for electrons is about 1.5 sr.
Electron momentum resolution is a function of the scattered electron angle
and it varies from 0.5\% for $\theta \leq 30^{\circ}$ up to
1-2\% for $\theta > 30^{\circ}$. The angular resolution
is approximately constant and approaching 1 mrad for polar and 4 mrad
for azimuthal angles: the resolution on the momentum transfer
ranges therefore from 0.2 up to 0.5 \%. The missing mass
resolution was estimated 2.5 MeV for beam energy less than 3 GeV and
about 7 MeV for larger energies.
To study all possible multi-particle production, the acquisition trigger
was configured to require at least one electron candidate in any of
the sectors, where an electron candidate was defined as
the coincidence of a signal in the EC and Cherenkov modules
for each sector separately.

The accumulated statistics at the five energies is large enough ($> 6 \times 10^8$
triggers)
to allow the extraction of the inclusive cross section with a rather small
statistical error ($\le 5$\%), in small $x$ and $Q^2$ bins ($\Delta x=0.009$,
$\Delta Q^2=0.05$ GeV$^2$).  The determination of the
systematic error was more critical. CLAS is a large acceptance spectrometer and the response
depends on the energy $E^{\prime}$ and the angle $\theta$ of the scattered
electron.
Determining the systematic effects of these, and other experimental
parameters, is both necessary and complex.
Consequently,
we dedicate the next subsections to the discussion of the analysis procedure.

\subsection{Momentum Correction}
Determining the momentum of a charged particle measured with CLAS depends
on a proper understanding of the magnetic field geometry. Due to the
complexity of the detector and particularly the torus magnet system,
it is crucial to check the reliability of the momentum determined by the DC
tracking system. For this reason the position of the elastic peak was
extracted from the measured inclusive electron cross
section and compared to the theoretical value. A correction to the
scattered electron momentum was applied to shift the elastic peak to the
accepted value. The momentum correction obtained was small
(from 2 to 7 MeV in $W$, depending on the beam energy) and
resulted in significant improvement in the width of elastic peak .

The systematic error on the correction was estimated by
comparing the position of the well known second resonance
($S_{11}(1535)+D_{13}(1520)$) to the position given in Ref.~\cite{Bodek,Stein}.
The position difference $\Delta W$ affects the cross section evaluation.
The relative systematic error on the momentum correction is
therefore given by
\begin{equation}\label{eq:mom_corr}
\delta_{mom}(x,Q^2)=\frac{|\sigma^B(W-\frac{\Delta W}{2},Q^2)-
\sigma^B(W+\frac{\Delta W}{2},Q^2)|}{\sigma^B(W,Q^2)}~,~~~~
\end{equation}
\noindent where $\sigma^B$ represents the Bodek fit according to the parametrization
from Ref.~\cite{Bodek,Stein}.  The systematic error calculated with Eq.~\ref{eq:mom_corr}
is given in Table.~\ref{table:syserr}.

\subsection{Electron Identification and Pion Rejection}\label{sec:ElPid}
The electrons were identified by a combined off-line analysis of the
signals from the four detector systems (DC, TOF, CC and EC). Only those
electron candidates that were detected inside the most uniform
(``fiducial'') detector volume were analyzed. The electron yield was
corrected separately in each kinematic bin, for pion contamination,
detection efficiency and radiative corrections.
\begin{figure}[!h]
\includegraphics[bb=1cm 6cm 20cm 23cm, scale=0.4]{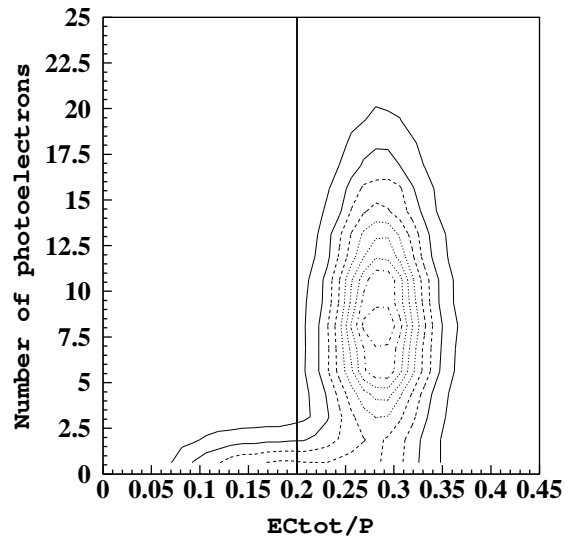}
\caption{\label{fig:PhEvsECI} Photoelectron distribution in the Cherenkov
detector versus the energy deposited in the EC detector divided by the momentum of
the particle as determined by the drift chamber.
The black vertical line represents the cut to reduce the pion
contamination.}
\end{figure}

\begin{figure}[!h]
\includegraphics[bb=1cm 6cm 20cm 23cm, scale=0.4]{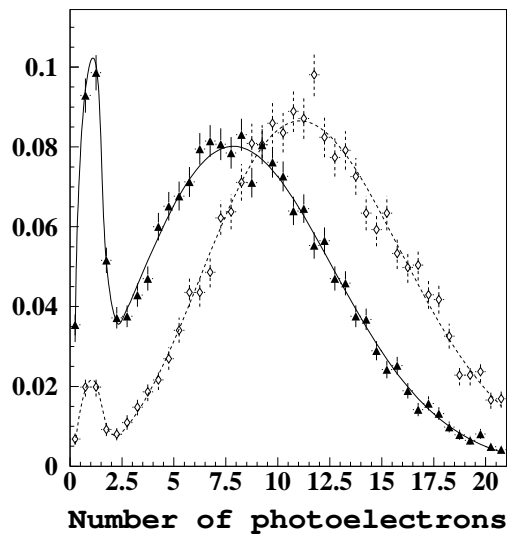}
\caption{\label{fig:FitPhe} The fitted photoelectron distribution
for two different sets of kinematics after removing some
of the pion contamination via the $EC_{tot}/P$ cut:
solid triangles show the distribution obtained with $4$ GeV beam,
scattered electron angle $\theta=31^{\circ}$ and momentum $P=1$ GeV;
empty diamonds represent data taken with $2.5$ GeV beam,
scattered electron angle $\theta=41^{\circ}$ and momentum $P=1$ GeV.}
\end{figure}

The photoelectron distribution in the CC depends on the
kinematics, and the contaminating pion peak can be completely removed only
with large efficiency losses of about 30\% in several kinematic regions.
Therefore the pion contamination was removed by a two-step procedure.
Electrons producing a large number of photoelectrons (see Fig.~\ref{fig:FitPhe})
were identified by an energy cut in the EC detector response.  The
pion contamination to electrons producing a small number of photoelectrons
was removed by analysis of the CC response.

As an example of the first step, Fig.~\ref{fig:PhEvsECI} shows the CC photoelectron
distribution $N_{phe}$ as a function of the fraction of energy deposited in the EC
detector $EC_{tot}/P$ for negatively charged particles emitted at
$\theta < 35^{\circ}$,
momentum $P<1$~GeV/c, and a beam energy of $2.5$~GeV. The regions
corresponding to pions ($N_{phe} \leq 2.5$) and to electrons
($EC_{tot}/P \geq 0.25$) cannot be clearly separated and only the
pion contamination to the left of the solid line can be removed without
affecting the electron detection efficiency.
The remaining pion contamination and the correction of the Cherenkov
efficiency for electrons $F_{phe}$ have determined
by a combined fit of the measured photoelectron distribution with two
Poisson distributions convoluted with a Gaussian function to account for
the finite photomultiplier resolution as shown in Fig.~\ref{fig:FitPhe}.

The fit was performed separately for each sector over the whole kinematics data set
($\theta=20^{\circ} - 50^{\circ}$ and $W=0.9 - 2.5$~GeV)

To minimize the errors, the fit was performed in rather large bins
($\Delta\theta=2^{\circ}$ and $\Delta E^{\prime} \sim 0.1$~GeV).
Therefore, in order to apply the correction to the measured cross
section, which was obtained with smaller bins, values of the correction
were parametrized with the polynomial function:
\begin{equation}
F_{phe}(\nu,\theta)=1+A(\nu-\nu_0)+(B+C\theta)(\nu-\nu_0)^2~,
\label{eqn:d_pc4}
\end{equation}
\noindent
where $A$, $B$, $C$ and $\nu_0$ are free parameters and $\nu=E_0-E^{\prime}$
the electron energy transfer. The related systematic error
$\delta_{phe}=\delta{F_{phe}}/F_{phe}$ is mainly due to the low
statistics in those bins corresponding to large $Q^2$ values and was found
to be less than $2$\%.

\subsection{Background Subtraction}
Since the pair production background has not been measured, its
contribution was estimated according to a model~\cite{BostedPP}
based on the Wiser fit~\cite{Wiser} of the inclusive pion photo-production reaction. The most
important source of $e^+e^-$ pairs in the CLAS is due to $\pi^0$ production,
which either decays to $\gamma e^+e^-$ (Dalitz decay) or to $\gamma \gamma$,
with subsequent photon conversion to $e^+e^-$. The model was carefully
checked, and it was in good agreement with the measured positron
cross sections~\cite{BostedPP};  the difference was always less than 30\%.
The value of the correction was assumed to be equal to the ratio of the inclusive
$e^+$ production cross section $\sigma_{e^+}$ over Bodek's fit~\cite{Bodek,Stein}
including radiative processes (tail from the elastic peak, bremsstrahlung, and
Schwinger correction)
$\sigma^B_{rad}$:
\begin{equation}
F_{e^+e^-}(E_0,E^{\prime},\theta)=
\frac{1}{1+\frac{\sigma_{e^+}(E_0,E^{\prime},\theta)}
{\sigma^B_{rad}(E_0,E^{\prime},\theta)}}~.
\label{eqn:d_pp1}
\end{equation}

The correction is generally small as expected in Ref.~\cite{f2-hc},
therefore it was applied only for $E_0 > 2.0$~GeV and $W>1.7$~GeV,
where it was about 2\%.
The relative systematic error
$\delta_{e^+e^-}=\delta{F_{e^+e^-}}/F_{e^+e^-}$ from this correction
was estimated using uncertainties on $\sigma_{e^+}$ given in Ref.~\cite{BostedPP}.

In order to remove the contribution of scattering on the target walls,
the empty target data were analyzed in
the same way and subtracted from the inclusive data, after proper
normalization.
An additional source of background originating from knock-on electrons
produced in the supporting structure of the detector was estimated and
it found to be smaller than 0.3\%.

\subsection{Simulations}
Due to the complexity of the CLAS detector the only way to study its
response functions is to perform complete computer simulations,
describing each subsystem in detail including all
materials that make up each detector.
The simulations
of detector response to the scattered electron
were performed according to the following procedure:

\begin{enumerate}
\item 
Electron scattering events
were generated by a random event generator with the
probability distributed according to $\sigma^B_{rad}$, described above.
The values of
elastic and inelastic cross sections
of the electron-proton scattering
were taken from existing fits of
world data, in references~\cite{Bosted} and~\cite{Bodek,Stein}, respectively. The internal
radiative processes contribution was added according to
calculations~\cite{Mo}.

\item The generated events were passed through the standard
CLAS GEANT-based simulation program~\cite{GSIM}, to model the detector response.

\item The results of the previous stage were further processed to make
the detector response more realistic by adding the effects of electronic
noise, background, dead wires and scintillator paddles.

\item Finally, the efficiency was calculated in each kinematic bin as a
ratio of the number of reconstructed events $N_{rec}$ over the number of
generated events $N_{gen}$:
\begin{equation}
\epsilon_{eff}(E_0,x,Q^2)=
\frac{N_{rec}(E_0,x,Q^2)}{N_{gen}(E_0,x,Q^2)}~.
\label{eq:d_e_s1}
\end{equation}
\end{enumerate}
\noindent The
electron detection
efficiency obtained from simulations is about 97\% and
approximately constant inside the fiducial region of the detector over
the whole available kinematics.

\begin{figure}
\includegraphics[bb=1cm 6cm 20cm 23cm, scale=0.4]{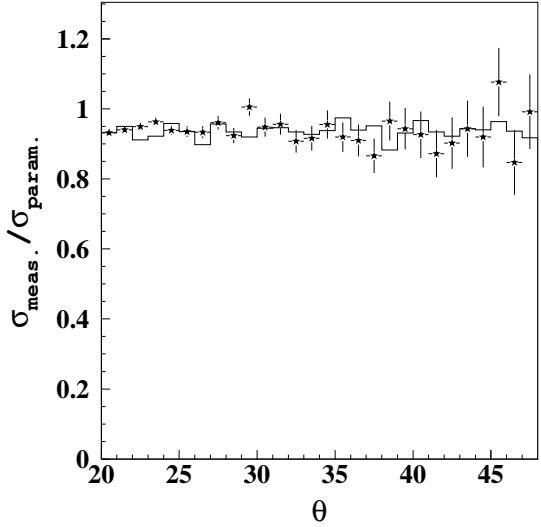}
\caption{\label{fig:SimEl} Typical ratio of the measured elastic
scattering cross section to the parametrization from Refs.~\cite{Bosted,hall-a}
(points) with radiative corrections, in comparison to
that obtained from the simulations (continuous line); errors are statistical
only.}
\end{figure}

In order to
test the reliability of the simulation procedure and
to check a proper absolute normalization,
the well known elastic
scattering cross section was extracted from the same data set
($\frac{d\sigma}{d\Omega}_{exp}$) and compared to the simulated cross section
($\frac{d\sigma}{d\Omega}_{sim}$). The two cross sections are in good
agreement within statistical and systematic errors as shown in
Fig.~\ref{fig:SimEl}.

The relative systematic deviation of the elastic cross section obtained from
simulations and from these data $\delta_{eff}$, was calculated for each
beam energy and scattered electron angle
(in bins of one degree on the accessible interval from 20 to 50 degrees)
according to:
\begin{eqnarray}\nonumber
\delta^2_{eff}(E_0,\theta)&+&\delta^2_{exp}(E_0,\theta)=\\
&&\Biggl[\frac{
\frac{d\sigma}{d\Omega}_{exp}(E_0,\theta)-
\frac{d\sigma}{d\Omega}_{sim}(E_0,\theta)}
{\frac{d\sigma}{d\Omega}_{fit}(E_0,\theta)}\Biggr]^2~,~~~~~~
\label{eq:d_e_ss1}
\end{eqnarray}
\noindent
where $\frac{d\sigma}{d\Omega}_{fit}$ is the parametrization described
in Ref.~\cite{Bosted} and $\delta_{exp}$ is the statistical error of
the measured elastic cross section. For the error propagation $\delta_{eff}$
was parametrized by a linear function of the scattered electron
angle $\theta$.

\subsection{Inclusive Inelastic Cross Section}\label{sec:d_ii}
Since the Monte Carlo simulations were shown to be reliable, they were used to evaluate
efficiency, acceptance, bin centering and radiative corrections.
For each kinematic bin, the inclusive cross
section $d\sigma$ and the structure function $F_2$ were extracted
directly from the raw electron yield $N_{exp}$ normalized to the
integrated luminosity 
and corrected for efficiency, acceptance, bin centering,
and radiative effects as follows:
\begin{eqnarray}
\frac{d^2 \sigma}{d x dQ^2} & = &
\frac{1}{\rho \frac{N_A}{M_A} L Q_{tot}} \frac{N_{exp}(x,Q^2)}
{\epsilon(x,Q^2)}
 \nonumber\\
&&F_{phe}(x,Q^2)
F_{e^+e^-}(x,Q^2)~,
\label{eq:d_ii2}
\end{eqnarray}
\noindent
where $\rho$ is the density of liquid $H_2$ in the target, $N_A$ is
the Avogadro constant, $M_A$ is the target molar mass, $L$ is the
target length, $Q_{tot}$ is the total charge in the Faraday Cup (FC) and $\epsilon(x,Q^2)$
is the efficiency defined in Eq.~\ref{eq:d_e_s1} with the radiative
and bin-centering correction factors according to:
\begin{equation}
\epsilon(x,Q^2)=\epsilon_{eff}(x,Q^2)\epsilon_{rad}(x,Q^2)
\epsilon_{bin}(x,Q^2)~,
\label{eq:d_ii3}
\end{equation}
\noindent
where
\begin{equation}
\epsilon_{rad}=\frac{\sigma^B_{rad}}{\sigma^B}~~~
\mbox{and}~~~\epsilon_{bin}
=\frac{\int_{\Delta \tau} d\sigma^B}{\sigma^B}~,
\end{equation}
\noindent
and the integral was taken over the current bin area $\Delta \tau$.
The radiative correction factor $\epsilon_{rad}$ strongly varies
in the explored kinematic range from 0.85 up to 1.6. Fortunately,
the largest correction was contributed by the elastic peak
tail for which calculations are very accurate (see Refs.~\cite{Mo,Akushevich}).

All systematic uncertainties were propagated in quadrature to the
final relative systematic error:
\begin{eqnarray}
&&\delta_{sys} (x,Q^2) = 
\bigl[\delta_{eff}^2(x,Q^2)+\delta_{phe}^2(x,Q^2)+ \nonumber\\
&&\delta_{e^+e^-}^2(x,Q^2)+\delta_{mom}^2(x,Q^2)+
\delta_{rad}^2(x,Q^2)\bigr]^{1/2}~,~
\label{eq:d_se1}
\end{eqnarray}
\noindent
where $\delta_{rad}$ is the systematic uncertainty on the radiative
correction, given by:
\begin{equation}
\delta_{rad}(E,x,Q^2)=\Bigl|\epsilon^{TSAI}_{rad}(E,x,Q^2)-
\epsilon^{POLRAD}_{rad}(E,x,Q^2)\Bigr|~,
\label{eq:d_rc_in11}
\end{equation}
\noindent
where $\epsilon^{TSAI}_{rad}(E,x,Q^2)$ and
$\epsilon^{POLRAD}_{rad}(E,x,Q^2)$ are the radiative correction factors
in $\sigma^B_{rad}$ evaluated with two different approaches
(\cite{Mo},~\cite{Akushevich}).
These two approaches use different parametrizations of the elastic
(\cite{Bosted} and \cite{Bilenka})
and inelastic (\cite{Bodek,Stein} and~\cite{Brasse}) cross sections as well as
different calculation techniques.

$\delta_{rad}(E,x,Q^2)$
varies in the kinematic range of the experiment from 0 to 20\% while
the average value is 3\%.
A minimum radiative correction systematic error of 1.5\%
was assumed.

\subsection{Structure Function $F_2(x,Q^2)$}\label{sec:d_sf}
The structure function $F_2(x,Q^2)$ was extracted from the inelastic
cross section using the fit of the function
$R(x,Q^2)\equiv \sigma_L / \sigma_T$ developed in~\cite{Ricco1} and
described in Appendix~\ref{appendix1}.
The inclusive electron scattering cross section can
be expressed in terms of the well known structure functions
$W_1$ and $W_2$ as~\cite{Roberts}:
\begin{equation}
\frac{d^2\sigma}{d\Omega dE^{\prime}}=
\sigma_{Mott}\bigl\{2W_1(x,Q^2) \tan^2{\frac{\theta}{2}} +W_2(x,Q^2)\bigr\}~,
\end{equation}
\noindent
where the Mott cross section is given by:
\begin{equation}
\sigma_{Mott}=\frac{\alpha^2 \cos^2{\frac{\theta}{2}}}{4E^2\sin^4{\frac{\theta}{2}}}~.
\end{equation}
\noindent Therefore, the structure function $F_2=\nu W_2$ is given by:
\begin{equation}
F_2(x,Q^2)=\frac{1}{\sigma_{Mott}}\frac{d^2\sigma}
{dx dQ^2}J
\frac{\nu}{1+\frac{1-\epsilon}{\epsilon}\frac{1}{1+R}}~,
\label{eq:d_sf1}
\end{equation}
\noindent
where $J$ is the Jacobian given by
\begin{equation}
J=\frac{x(s-M^2)}{2\pi M\nu}E^{\prime}~,
\end{equation}
\noindent
where $\it s$ is the squared invariant mass of the initial electron-proton system $s=M^2+2EM$ and
$\epsilon$ is the polarization parameter defined as
\begin{equation}
\epsilon \equiv \Biggl(
1+2\frac{\nu^2+Q^2}{Q^2} \tan^2{{\theta \over 2}}
\Biggr )^{-1}~.
\label{eq:d_sf2}
\end{equation}

The function $R(x,Q^2)$ is poorly known in the
resonance region, however the structure function
$F_2$ in the relevant kinematic range is very insensitive
to the value of $R$. In fact even a 100\% systematic uncertainty
on $R$ gives only a few percent uncertainty on $F_2$.
The relative total systematic error is given by:
\begin{equation}
\delta^{sys}_{F_2}(x,Q^2)=
\Biggl[
\delta_{sys}^2(x,Q^2)+
\Biggl(\frac{1-\epsilon}{1+\epsilon R}
\frac{\delta_R}{1+R} \Biggr )^2
\Biggr]^{1/2}~.
\label{eq:d_sf6}
\end{equation}
\noindent The uncertainties of $R$ given in Ref.~\cite{Ricco1} were propagated
to the resulting $F_2$, and the actual systematic errors introduced by
$\delta_R$ were always lower than 3\%.

\begin{figure}
\includegraphics[bb=1cm 6cm 20cm 23cm, scale=0.4]{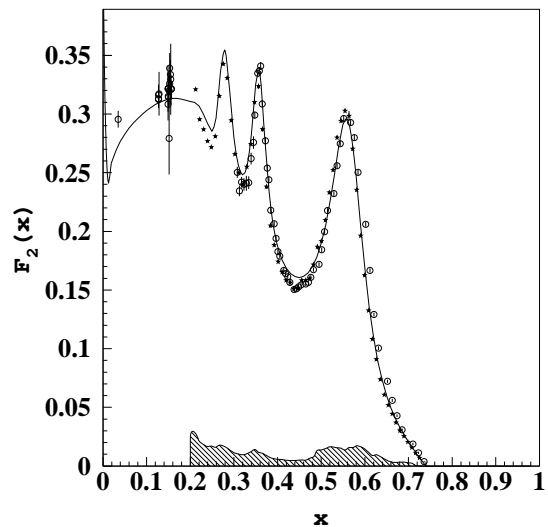}
\caption{\label{fig:f2comp} Structure function $F_2(x,Q^2)$ at
$Q^2=0.775$ GeV$^2$: stars represent experimental data obtained in
the present analysis
with systematic errors indicated by the hatched area,
empty circles show data from previous experiments
~\cite{f2-hc,BCDMS,E665,NMC,H1a,H1b,H1c,H1d,H1e,ZEUSa,
ZEUSb,ZEUSc,ZEUSd,ZEUSe,ZEUSf,SLAC}
and the solid line represents
the parametrization from Ref.~\cite{Ricco1}.
}
\end{figure}

The combined statistical and systematic
precision of the obtained structure function $F_2$
is strongly dependent on kinematics and the statistical errors
vary from 0.2\% up to 30\% at the largest $Q^2$ where
statistics are very limited.
Fig.~\ref{fig:f2comp} shows a comparison between the $F_2$ data
from CLAS and the other world data in the $Q^2=0.775$ GeV$^2$ bin.
The observed discrepancies with the data from Ref.~\cite{f2-hc}
which fill the large $x$ region in Fig.~\ref{fig:f2comp}
are mostly within the systematic errors. Because
of the much smaller bin centering corrections in this $Q^2$ region
our data are in a better agreement with data previously measured at SLAC,
given in Ref.~\cite{Stein},
and the parameterization of those from Ref.~\cite{Bodek,Stein}.
The average statistical uncertainty is about 5\%;
the systematic uncertainties range from 2.5\% up to 30\%, with
the mean value estimated as 7.7\% (see Table~\ref{table:syserr}).
The values of $F_2(x,Q^2)$ determined using our data are tabulated
elsewhere \cite{CLAS_note}.

\begin{table}[!h]
\begin{center}
\caption{Range and average of systematic errors on $F_2$.}
\label{table:syserr}
\vspace{2mm}
\begin{tabular}{|c|c|c|} \hline
     Source of uncertainties                 & Variation range   & Average \\ \cline{2-3}
                                             & [\%]              & [\%]    \\ \hline
Efficiency evaluation                        & 1-9               & 4.3  \\ \hline
$e^+e^-$ pair production correction          & 0-3               & 0.3  \\ \hline
Photoelectron correction                     & 0.1-2.2           & 0.6  \\ \hline
Radiative correction                         & 1.5-20            & 3.2  \\ \hline
Momentum correction                          & 0.1-30            & 3.5  \\ \hline
Uncertainty of $R=\frac{\sigma_L}{\sigma_T}$ & 0.5-5             & 2.4  \\ \hline
Total                                        & 2.5-30            & 7.7  \\ \hline
\end{tabular}
\end{center}
\end{table}

\subsection{Moments of the Structure Function $F_2$}\label{sec:d_nm}
As discussed in the introduction, the final goal of this analysis
is the evaluation of the Nachtmann moments of the structure
function $F_2$. The total Nachtmann moments were computed as the sum of
the elastic and inelastic moments:
\begin{equation}
M_n = M^{el}_n + M^{in}_n~.
\label{eq:d_nm_t1}
\end{equation}

The contribution originating from the elastic peak was calculated
according to the following expression from Ref.~\cite{Ricco1}:
\begin{eqnarray}
M^{el}_n = \biggl (\frac{2}{1+r}\biggr )^{n+1}
&&\frac{3+3(n+1)r+n(n+2)r^2}{(n+2)(n+3)} \nonumber \\
&&\frac{G_E^2(Q^2)+\frac{Q^2}{4M^2} G_M^2(Q^2)}{1+\frac{Q^2}{4M^2}}~,
\label{eq:d_em1}
\end{eqnarray}
\noindent
where the proton form factors $G_E^2(Q^2)$ and $G_M^2(Q^2)$
are from Ref.~\cite{Bosted} modified according the recently measured
data on $G_E/G_M$~\cite{hall-a}, as described in Ref.~\cite{CLAS_note}.

The evaluation of the inelastic moment $M^{in}_n$ involves the
computation at fixed $Q^2$ of an
integral over $x$. For this purpose, in
addition to the results obtained from the CLAS data,
world data on the structure function $F_2$
from Refs.~\cite{f2-hc,BCDMS,E665,NMC,H1a,H1b,H1c,H1d,H1e,ZEUSa,
ZEUSb,ZEUSc,ZEUSd,ZEUSe,ZEUSf,SLAC} and
data on the inelastic cross section~\cite{csworld,Bodek,Stein}
were used to reach an adequate coverage (see Fig.~\ref{fig:xandQ2Domain}).
The integral over $x$ was performed numerically using the standard
trapezoidal method TRAPER~\cite{cernlib}.
Data from Ref.~\cite{EMC} were not included in the analysis due
to their inconsistency with other data sets as explained in detail
in Ref.~\cite{disfit2}, and data from Ref.~\cite{WA25,new-inclusive} were not
included due to the large experimental uncertainties.

\begin{figure}
\includegraphics[bb=1cm 6cm 20cm 23cm, scale=0.4]{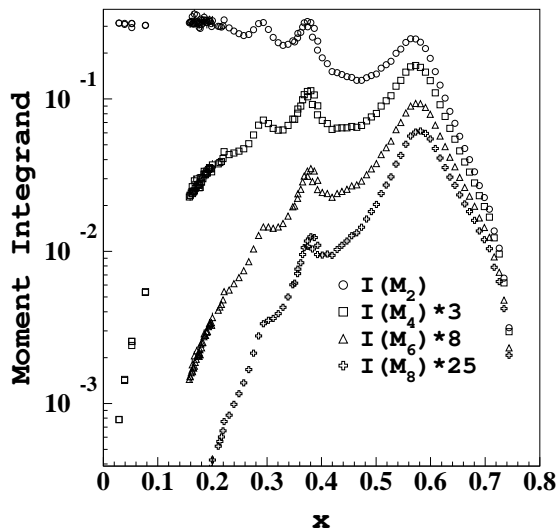}
\caption{\label{fig:intgr} Integrands of the inelastic Nachtmann moments at $Q^2=0.825$ GeV$^2$:
circles represent the integrand of the $M_2$;
squares show the integrand of the $M_4$;
triangles - the integrand of the $M_6$;
crosses - the integrand of the $M_8$.}
\end{figure}

The $Q^2$-range from 0.05 to 3.75 (GeV/c)$^2$ was divided
into $\Delta Q^2 = 0.05$~(GeV/c)$^2$
bins. Then within each $Q^2$ bin the world data were
shifted to the central bin value $Q^2_0$, using the fit of $F_2^B(x,Q^2)$ from
Ref.~\cite{Ricco1}. Here the fit $F_2^B(x,Q^2)$ consists of two parts,
a parametrization~\cite{Bodek,Stein} in the resonance region ($W<2.5$ GeV),
and a QCD-like fit from Ref.~\cite{disfit1} in the DIS ($W>2.5$ GeV):
\begin{equation}
F_2(x,Q^2_0)=\frac{F_2(x,Q^2)}
{F_2^{B}(x,Q^2)}F_2^{B}(x,Q^2_0)~.
\label{eq:d_nm2}
\end{equation}
\noindent
The difference between the real and bin-centered data
\begin{equation}
\delta^{cent}_{F_2}(x,Q^2)=F_2(x,Q^2)\Biggl|1-
\frac{F^B_2(x,Q^2_0)}{F_2^B(x,Q^2)}\Biggr|~~,
\end{equation}
\noindent
was added to the systematic errors of $F_2$ in extracting
the Nachtmann moments.
As an example, Fig.~\ref{fig:intgr} shows the integrands of the first four moments
as a function of $x$ at fixed $Q^2$.  The significance of the large $x$ region for
various moments can clearly be seen .

To have a data set dense in $x$, which reduces the error in
the numerical integration, we performed an interpolation, at each
fixed $Q^2_0$, when two contiguous
experimental data points differed by more than $\nabla$. The value of
$\nabla$ depended on kinematics;
in the resonance region where the structure function exhibits
strong variations, $\nabla$ had to be smaller than half the
resonance widths and was parametrized as
$\nabla =0.04/\Bigl[1+\sqrt{Q^2/10}\Bigr]$. Above the resonances, where
$F_2$ is smooth, we only accounted for the fact that the available $x$ region
decreases with decreasing $Q^2$ ($\nabla =0.1\Bigl[1+\sqrt{Q^2/10}\Bigr]$).
Finally in the low $x$ region ($x<0.03$) where the $F_2$ shape
depends weakly on $Q^2$, but strongly on $x$, we set $\nabla =0.015$.
Changing these $\nabla$ values by as much as a factor two
produced changes in the moments that were
much smaller than the systematic errors.

To fill the gap within two contiguous points $x_a$ and $x_b$, we
used the interpolation function $F_2^{int}(x,Q^2_0)$ defined as
the parametrization from Ref.~\cite{Ricco1} normalized to the
experimental data on both edges of the interpolating range.
Assuming that the shape of the fit is correct:
\begin{equation}
F_2^{int}(x,Q^2_0)=\alpha(Q^2_0)F_2^{B}(x,Q^2_0)~,
\label{eq:d_nm_i1}
\end{equation}
\noindent
where the normalization factor $\alpha(Q^2_0)$ is defined as the
weighted average, evaluated using all experimental points located
within an interval $\Delta$ around $x_a$ or $x_b$:
\begin{eqnarray}
&&\alpha(Q^2_0)=\delta_N^2(Q^2_0)\Biggl [
\sum\limits_{i}^{|x_i-x_a|<\Delta}
\frac{F_2(x_i,Q^2_0)/F_2^{B}(x_i,Q^2_0)}{\bigl(\delta_{F_2}^{stat}
(x_i,Q^2_0)\bigr)^2}+ \nonumber \\
&&\sum\limits_{j}^{|x_j-x_b|<\Delta}
\frac{F_2(x_j,Q^2_0)/F_2^{B}(x_j,Q^2_0)}{\bigl(\delta_{F_2}^{stat}
(x_j,Q^2_0)\bigr)^2}\Biggr ]~,
\label{eq:d_nm_i2}
\end{eqnarray}
\noindent
where $\delta_{F_2}^{stat}(x_j,Q^2_0)$ is the statistical error
relative to $F_2^{B}$ and
\begin{eqnarray}
\delta_N(Q^2_0)=\Biggl [
\sum\limits_{i}^{|x_i-x_a|<\Delta}\frac{1}{\bigl(\delta_{F_2}^{stat}
(x_i,Q^2_0)\bigr)^2}+ \nonumber \\
\sum\limits_{j}^{|x_j-x_b|<\Delta}\frac{1}{\bigl(\delta_{F_2}^{stat}
(x_j,Q^2_0)\bigr)^2}
\Biggr ]^{-1/2}
~,
\label{eq:d_nm_i3}
\end{eqnarray}
\noindent
is the statistical uncertainty of the normalization.
Therefore, the statistical error of the moments calculated according
the trapezoidal rule~\cite{cernlib} was increased
by adding the linearly correlated contribution from each interpolation
interval as follows:
\begin{eqnarray}
\delta^{norm}_n (Q^2_0) = &\delta_N& (Q^2_0)
\int\limits_{x_a}^{x_b}
dx \frac{\xi^{n+1}}{x^3}F_2^{B}(x,Q^2_0) \nonumber \\
&& \frac{3+3(n+1)r+n(n+2)r^2}{(n+2)(n+3)}~.
\label{eq:d_nm_i25}
\end{eqnarray}

\begin{figure}
\includegraphics[bb=1cm 6cm 20cm 23cm, scale=0.4]{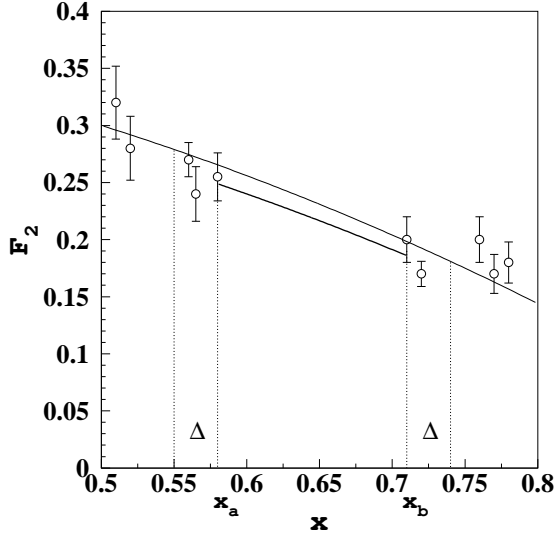}
\caption{\label{fig:interp} Example of the interpolating
procedure. The meaning of the curves and symbols is described in the text.}
\end{figure}

Since we average the ratio
$F_2(x_i,Q^2_0)/F_2^{B}(x_i,Q^2_0)$, $\Delta$ is not affected
by the resonance structures, and its value was fixed to have
more than two experimental points
in most cases; therefore,
$\Delta$ was chosen equal to 0.03 in the resonance and in the very
low $x$ regions, and to 0.05 in the DIS region. In Fig.~\ref{fig:interp}
we show how this interpolation is applied: the thin continuous line
represents the original function $F_2^{B}(x,Q^2)$ and the heavy
continuous line represents the result of the interpolation
$F_2^{int}(x,Q^2)$. We also checked that the moments do not
show any dependence on the $\Delta$ values.   

To fill the gap between the last experimental point and one of
the integration limits ($x_a=0$ or $x_b=1$) we performed an
extrapolation at each fixed $Q^2_0$ using $F_2^{B}(x,Q^2_0)$
including its systematic error given in
Ref.~\cite{Ricco1}.

As an extension of the analysis, the world data at $Q^2$ above
$5$~(GeV/c)$^2$ were analyzed in the same way as described above.
The only differences were the $Q^2$ bin size, which was chosen
equal to 5\% of $Q^2$, and the values of the parameters $\nabla$ and
$\Delta$. In addition, the bins were situated
not continuously, but only where data exist.
Since at large $Q^2$ the shape of $F_2(x)$
remains almost constant with changing $Q^2$ both parameters
$\nabla$ and $\Delta$ were fixed:
in the resonance region ($W<1.8$~GeV) to value $0.01$;
$0.1$ in the DIS;
and $\nabla=0.005$ and $\Delta=0.01$ at very low $x$ ($x<0.03$).
The results for $M^{in}_n(Q^2)$ did not exhibit any significant
dependence on the choice of the parameter values.
The results are reported together with their statistical and
systematic errors in Table~\ref{table:r_nm1}.

\subsection{Systematic Errors of the Moments}\label{sec:d_nm_ds}
The systematic error consists of genuine uncertainties in the
data given in Refs.~\cite{f2-hc,BCDMS,E665,NMC,H1a,H1b,H1c,H1d,H1e,ZEUSa,
ZEUSb,ZEUSc,ZEUSd,ZEUSe,ZEUSf,SLAC,Bodek,Stein,csworld} and uncertainties in
the evaluation procedure. To estimate the first type of error
we had to account for using many data sets
measured, in different laboratories, and with different detectors.
In the present analysis we assume that
different experiments are independent and therefore
only systematic errors within one data set are correlated.

Thus, an upper limit for the contribution
of the systematic error from each data set was evaluated in the
following way: 
\begin{itemize}
\item we first applied a simultaneous shift to all experimental
points in this set by an amount equal to their systematic error;
\item then the inelastic $n$-th moments obtained using these
distorted data $\tilde{M}^{in}_{n(i)}(Q^2)$ were compared to the
original moments $M^{in}_n(Q^2)$ evaluated with no systematic shifts;
\item finally the deviations for each data set were summed
in quadrature as independent values:
\begin{equation}
\delta_n^D(Q^2) = \frac{1}{M^{in}_n(Q^2)}\sqrt{\sum\limits_i^{N_{S}}
\Bigl(\tilde{M}^{in}_{n(i)}(Q^2)-M^{in}_n(Q^2)\Bigr)^2}~,
\label{eq:d_nm_ds1}
\end{equation}
\noindent
where $N_{S}$ is the number of available data sets.
The resulting error was summed in quadrature to $\delta_n^{norm}(Q^2)$
to finally evaluate the total systematic error of the $n$-th moment.
\end{itemize}

The second type of error is related to the bin centering, interpolation and
extrapolation. The bin centering systematic uncertainty was
estimated as follows:
\begin{equation}
\delta^C_n(Q^2)=\sum_i K_n(x_i,Q^2) w_i(Q^2) \delta^{cent}_{F_2}(x_i,Q^2)~,
\end{equation}
\noindent
where according the Nachtmann moment definition and the trapezoidal
integration rule:
\begin{eqnarray}
K_n(x_i,Q^2) &=&
\frac{\xi_i^{n+1}}{x_i^3} \frac{3+3(n+1)r_i+n(n+2)r_i^2}{(n+2)(n+3)} \nonumber \\
w_i(Q^2) &=& (x_{i+1}-x_{i-1})/2~.
\end{eqnarray}

The relative systematic error of the interpolation
was estimated as the possible change of the fitting function slope
in the interpolation interval, and it was evaluated as a difference
in the normalization at different edges:
\begin{eqnarray}
\delta_S(Q^2_0)=\Biggl | \frac{1}{N_i}\sum\limits_{i}^{|x_i-x_a|<\Delta}
\frac{F_2(x_i,Q^2_0)}{F_2^{B}(x_i,Q^2_0)}- \nonumber \\
\frac{1}{N_j}\sum\limits_{j}^{|x_j-x_b|<\Delta}
\frac{F_2(x_j,Q^2_0)}{F_2^{B}(x_j,Q^2_0)} \Biggr |~,
\label{eq:d_nm_i4}
\end{eqnarray}
\noindent
where $N_i$ and $N_j$ are the number of points used to evaluate
the sums. Since the structure function $F_2(x,Q^2)$ is a very
smooth function of $x$ below resonances, on the limited $x$-interval
(smaller than $\nabla$) the linear approximation gives
a good estimate. Thus, the error given in Eq.~\ref{eq:d_nm_i4}
accounts for such a linear mismatch between the fitting
function and the data on the interpolation interval.
Meanwhile, the CLAS data cover all the resonance region and
no interpolation was used there.
Therefore, the total systematic error introduced in the corresponding
moment by the interpolation can be estimated as
\begin{eqnarray}
\delta^I_n (Q^2_0) = &\delta_S& (Q^2_0)
\int\limits_{x_a}^{x_b}
dx \frac{\xi^{n+1}}{x^3} F_2^{B}(x,Q^2_0) \nonumber \\
&& \frac{3+3(n+1)r+n(n+2)r^2}{(n+2)(n+3)}~.
\label{eq:d_nm_i5}
\end{eqnarray}

The systematic errors obtained by these procedures were summed in quadrature:
\begin{eqnarray}
\delta^P_n(Q^2)=\sqrt{(\delta^D_n(Q^2))^2+(\delta^C_n(Q^2))^2+(\delta^I_n(Q^2))^2}~.
\label{eq:d_nm_i55}
\end{eqnarray}

\begin{figure}
\includegraphics[bb=1cm 6cm 20cm 23cm, scale=0.4]{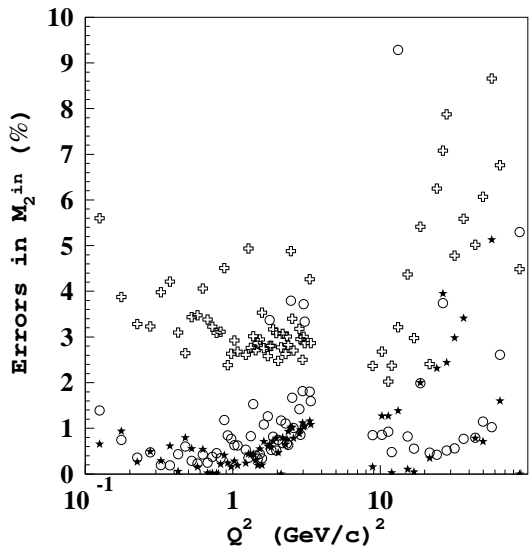}
\caption{\label{fig:SepErr} Errors of the inelastic Nachtmann moment $M_2$ in percentage:
empty circles represent statistical errors;
empty crosses show the systematic error obtained in Eq.~\ref{eq:d_nm_i55};
the difference between inelastic moments extracted using two different
$F_2$ parametrizations \cite{Ricco1} and \cite{Forte} at $W>2.5$ GeV is shown by stars.}
\end{figure}

In order to study the systematic error on the extrapolation at very low $x$
we have performed a test of the functional form dependence
comparing moments presented here with those obtained by using the fitting function from 
the neural network parametrization of Ref.~\cite{Forte}. The difference
is significant only for $M_2$ and it appeared to be smaller than $\delta^P_n(Q^2)$
given by Eq.~\ref{eq:d_nm_i55} (see Fig.~\ref{fig:SepErr}).
The difference was added to $\delta^P_n(Q^2)$ in quadrature to evaluate
the total systematic error of the $n$-th moment.

\section{Extraction of leading and higher twists}\label{sec:Discussion}
\begin{figure}
\includegraphics[bb=1cm 6cm 20cm 23cm, scale=0.4]{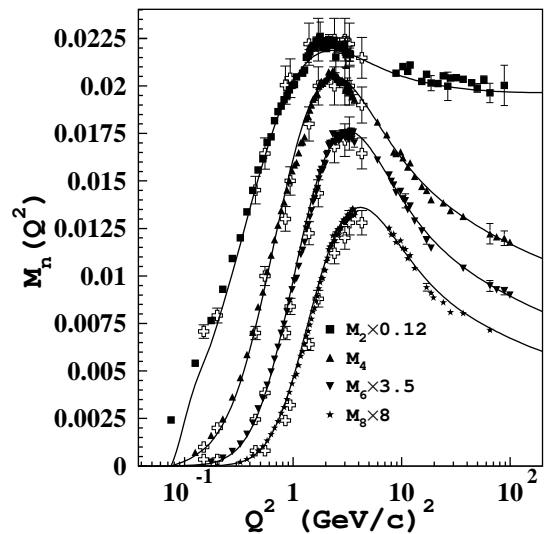}
\caption{\label{fig:NachtMom}
The inelastic Nachtman moments extracted from
the world data, including the new CLAS results, are shown
as the solid shapes
; while the solid line represents moments
obtained in Ref.~\cite{Ricco1}.
The empty crosses indicate the Nachtman
moments determined in Ref.~\cite{f2mom-hc}. Errors are statistical only.}
\end{figure}

In this Section, we present our twist analysis of the
moments we have extracted, which are presented in Fig.~\ref{fig:NachtMom}. As already shown in
Refs.~\cite{Ricco1} and~\cite{SIM00}, the extraction of
higher twists at large $x$ depends significantly on the
effects of 
pQCD high-order corrections.
In particular,
the use of the well established NLO approximation for the
leading twist is known to lead to unreliable results for the higher
twists~\cite{SIM00}. Therefore, hereafter we follow Ref.~\cite{SIM00},
where the pQCD corrections beyond the NLO are estimated according
to soft gluon resummation (SGR) techniques.

As far as power corrections are concerned, several higher-twist
operators exist and mix under the renormalization-group equations.
Such mixings are rather involved and the number of mixing
operators increases with the order $n$ of the moment. Since a complete
calculation of the higher-twist anomalous dimensions is not yet
available, we use the same phenomenological ansatz already adopted
in Refs.~\cite{Ricco1} and~\cite{SIM00}. Thus, our extracted
Nachtmann moments are analyzed in terms of the following twist expansion
\begin{eqnarray}
     M_n^N(Q^2) = \eta_n(Q^2) + HT_n(Q^2) ~ ,
    \label{eq:twists}
\end{eqnarray}
\noindent where $\eta_n(Q^2)$ is the leading twist moment and
$HT_n(Q^2)$ is the higher-twist contribution given by~\cite{Ji}

\begin{eqnarray}\nonumber
HT_n(Q^2)=&&
a_n^{(4)}\biggl[\frac{\alpha_s(Q^2)}{\alpha_s(\mu^2)}\biggr]^{
\gamma_n^{(4)}}\frac{\mu^2}{Q^2}\\
+&&a_n^{(6)}\biggl[\frac{\alpha_s(Q^2)}{\alpha_s(\mu^2)}\biggr]^{
\gamma_n^{(6)}}\frac{\mu^4}{Q^4} ~ ,
\label{eq:HT}
\end{eqnarray}

\noindent where the logarithmic pQCD evolution of the twist-$\tau$
contribution is accounted for by the term
$[\alpha_s(Q^2) / \alpha_s(\mu^2)]^{\gamma_n^{(\tau)}}$
(corresponding to the Wilson coefficient $E_{n \tau}(\mu,Q^2)$
in Eq.~\ref{eq:i_m1}) with an {\em effective} anomalous dimension
$\gamma_n^{(\tau)}$ and the parameter $a_n^{(\tau)}$ (equal to the
matrix element $O_{n \tau}(\mu)$ in Eq.~\ref{eq:i_m1}) represents
the overall strength of the twist-$\tau$ term at the
renormalization scale $Q^2 = \mu^2$.
In Eqs.~\ref{eq:twists}-\ref{eq:HT} four higher-twist parameters
appear, while the twist-2 moment $\eta_n(Q^2)$ is generally given
by the sum of a non-singlet and singlet terms, leading to
three unknown parameters, namely the values of the gluon,
non-singlet and singlet quark moments at the factorization scale
$Q^2 = \mu^2$. However,  the decoupling in the pQCD evolution
of the singlet quark and gluon densities at large $x$ allows one to
consider a pure non-singlet evolution for $n \geq 4$
(cf.~\cite{Ricco1}).
This means that we have only one twist-2 parameter for $n \geq 4$,
namely the value of the twist-2 moment $\eta_n(\mu^2)$ at the
factorization scale $Q^2 = \mu^2$. The resummation of soft gluons
does not introduce any further parameter in the description of the
leading twist. Explicitly, for $n \geq 4$ the  leading twist
moment $\eta_n(Q^2)$ is given by  
 \begin{eqnarray}
       \eta_n(Q^2) & = & A_n \left[ {\alpha_s(Q^2) \over 
       \alpha_s(\mu^2)} \right]^{\gamma_n^{NS}} \nonumber \\
       & & \left\{ \left[1 +  {\alpha_s(Q^2) \over 2 \pi}
       C_{DIS}^{(NLO)} \right]  e^{G_n(Q^2)} + 
       \right.  \nonumber \\
       & & \left. {\alpha_s(Q^2) \over 4 \pi} R_n^{NS} \right\} ~ , 
       \label{eq:SGR}
 \end{eqnarray}
\noindent where the quantities $\gamma_n^{NS}$, $C_{DIS}^{(NLO)}$
and $R_n^{NS} $ can be read off from Ref.~\cite{SIM00}. In
Eq.~\ref{eq:SGR} the function $G_n(Q^2)$ is the key quantity of
the soft gluon resummation. At next-to-leading-log it reads as 
\begin{eqnarray}
       G_n(Q^2) = \mbox{ln}(n) G_1(\lambda_n) + G_2(\lambda_n) +O[\alpha_s^k 
       \mbox{ln}^{k-1}(n)] ~ ,
       \label{eq:Gn}
\end{eqnarray}
where $\lambda_n \equiv \beta_0 \alpha_s(Q^2) \mbox{ln}(n) / 4\pi$ and
 \begin{eqnarray}
       G_1(\lambda) & = & C_F {4 \over \beta_0 \lambda} \left[ \lambda + (1 
       -  \lambda) \mbox{ln}(1 - \lambda) \right] ~ , \nonumber \\
       G_2(\lambda) & = & - C_F {4 \gamma_E + 3 \over \beta_0} \mbox{ln}(1 
       - \lambda) \nonumber \\
       & & - C_F {8 K \over \beta_0^2} \left[ \lambda + \mbox{ln}(1 
       - \lambda) \right] \nonumber \\
      & & + C_F {4  \beta_1 \over \beta_0^3} \left[ \lambda + \mbox{ln}(1 - 
      \lambda) + {1 \over 2} \mbox{ln}^2(1 - \lambda) \right] ~ ,
     \label{eq:G1G2}
 \end{eqnarray}
\noindent with $C_F \equiv (N_c^2-1)/(2N_c)$,
$k = N_c (67/18 - \pi^2 / 6) - 5 N_f / 9$,
$\beta_0 = 11 - 2 N_f / 3$, and $N_f$ being the number of active
flavors. Note that the function $G_2(\lambda)$ is divergent for
$\lambda \to 1$; this means that at large $n$ (i.e. large $x$)
the soft gluon resummation cannot be extended to arbitrarily low
values of $Q^2$. Therefore, for a safe use of present SGR
techniques we will work far from the above-mentioned divergences
by limiting our analyses of low-order moments ($n \leq 8$) to
$Q^2 \geq 1$~(GeV/c)$^2$

All the unknown parameters, namely the twist-2 parameter $A_n$ as well
as the higher-twist parameters
$a_n^{(4)}, \gamma_n^{(4)}, a_n^{(6)}, \gamma_n^{(6)}$, were
simultaneously determined from a $\chi^2$-minimization procedure
in the $Q^2$ range between $1$ and $100$~(GeV/c)$^2$. In such a
procedure only the statistical errors of the experimental moments
were taken into account, as well as the updated Particle Data Group value
$\alpha_s(M_Z^2) = 0.118$~\cite{PDG}, and a renormalization scale
equal to $\mu^2 = 10$~(GeV/c)$^2$. The uncertainties
of the various twist parameters were then obtained by adding the
systematic errors to the experimental moments and by repeating
the twist extraction procedure. The parameter values are
reported in Table~\ref{table:twist1}, where it can be seen
that: the leading twist
is determined with a few percent uncertainty,
while the precision of the extracted higher twists increases with $n$
reaching an overall 10\% for $n=6$ and $8$,
thanks to the remarkable quality of the CLAS data at large $x$.
Note that the leading twist is
directly extracted from the data, which means that no specific
functional shape of the parton distributions is assumed in
our analysis. The contribution of higher twists to $M_2$ was
too small to be extracted by the present procedure.

\begin{table*}
\caption{\label{table:r_nm1}The inelastic Nachtmann moments for
$n=2,4,6$ and $8$ evaluated in the interval
$0.05 \le~Q^2 \le 100$~(GeV/c)$^2$. The moments are labeled
with * when the contribution to the integral by the experimental data is between
50\% and 70\%; all other values were evaluated
with more than 70\% data coverage. The data are reported
together with the statistical and systematic errors.}
\begin{ruledtabular}
\begin{tabular}{|c|c|c|c|c|} \cline{1-5}
$Q^2~[$(GeV/c)$^2]$ & $M_2(Q^2)$x$10^{-1}$ & $M_4(Q^2)$x$10^{-2}$ & $M_6(Q^2)$x$10^{-3}$ & $M_8(Q^2)$x$10^{-3}$ \\ \cline{1-5}
 0.075 & 0.202 $\pm$ 0.002 $\pm$ 0.009 & 0.016 $\pm$ 0.0005 $\pm$ 0.001 & 0.0019 $\pm$ 0.0001 $\pm$ 0.0001 &                                    \\ \cline{1-5}
 0.125 & 0.451 $\pm$ 0.006 $\pm$ 0.025 & 0.072 $\pm$ 0.002 $\pm$ 0.004 & 0.017 $\pm$ 0.001 $\pm$ 0.001 &                                    \\ \cline{1-5}
 0.175 & 0.638 $\pm$ 0.005 $\pm$ 0.025 & 0.162 $\pm$ 0.002 $\pm$ 0.007 & 0.060 $\pm$ 0.001 $\pm$ 0.003 & 0.0025 $\pm$ 0.0001 $\pm$ 0.0001   \\ \cline{1-5}
 0.225 & 0.775 $\pm$ 0.003 $\pm$ 0.026 & 0.248 $\pm$ 0.001 $\pm$ 0.008 & 0.119 $\pm$ 0.001 $\pm$ 0.004 & 0.0064 $\pm$ 0.0001 $\pm$ 0.0002   \\ \cline{1-5}
 0.275 & 0.910 $\pm$ 0.004 $\pm$ 0.030 & 0.364 $\pm$ 0.002 $\pm$ 0.015 & 0.218 $\pm$ 0.002 $\pm$ 0.010 & 0.0146 $\pm$ 0.0001 $\pm$ 0.0007   \\ \cline{1-5}
 0.325 & 1.000 $\pm$ 0.002 $\pm$ 0.040 & 0.465 $\pm$ 0.0005 $\pm$ 0.026 & 0.328 $\pm$ 0.0005 $\pm$ 0.020 & 0.0259 $\pm$ 0.00005 $\pm$ 0.0017   \\ \cline{1-5}
 0.375 & 1.114 $\pm$ 0.002 $\pm$ 0.047 & 0.587 $\pm$ 0.0005 $\pm$ 0.033 & 0.478 $\pm$ 0.0005 $\pm$ 0.030 & 0.0439 $\pm$ 0.00005 $\pm$ 0.0029   \\ \cline{1-5}
 0.425 & 1.209 $\pm$ 0.005 $\pm$ 0.037 & 0.704 $\pm$ 0.001 $\pm$ 0.034 & 0.644 $\pm$ 0.001 $\pm$ 0.038 & 0.0670 $\pm$ 0.0001 $\pm$ 0.0043   \\ \cline{1-5}
 0.475 & 1.298 $\pm$ 0.008 $\pm$ 0.036 & 0.839 $\pm$ 0.003 $\pm$ 0.023 & 0.858 $\pm$ 0.003 $\pm$ 0.024 & 0.1002 $\pm$ 0.0004 $\pm$ 0.0030   \\ \cline{1-5}
 0.525 & 1.347 $\pm$ 0.004 $\pm$ 0.047 & 0.916 $\pm$ 0.003 $\pm$ 0.038 & 1.010 $\pm$ 0.005 $\pm$ 0.046 & 0.1279 $\pm$ 0.0008 $\pm$ 0.0062   \\ \cline{1-5}
 0.575 & 1.419 $\pm$ 0.003 $\pm$ 0.049 & 1.023 $\pm$ 0.002 $\pm$ 0.050 & 1.215 $\pm$ 0.002 $\pm$ 0.068 & 0.1660 $\pm$ 0.0003 $\pm$ 0.0101   \\ \cline{1-5}
 0.625 & 1.444 $\pm$ 0.006 $\pm$ 0.059 & 1.110 $\pm$ 0.003 $\pm$ 0.041 & 1.413 $\pm$ 0.005 $\pm$ 0.057 & 0.2079 $\pm$ 0.0009 $\pm$ 0.0090   \\ \cline{1-5}
 0.675 & 1.514 $\pm$ 0.004 $\pm$ 0.051 & 1.191 $\pm$ 0.001 $\pm$ 0.062 & 1.603 $\pm$ 0.002 $\pm$ 0.098 & 0.2507 $\pm$ 0.0005 $\pm$ 0.0168   \\ \cline{1-5}
 0.725 & 1.554 $\pm$ 0.006 $\pm$ 0.050 & 1.267 $\pm$ 0.001 $\pm$ 0.059 & 1.785 $\pm$ 0.002 $\pm$ 0.102 & 0.2946 $\pm$ 0.0004 $\pm$ 0.0190   \\ \cline{1-5}
 0.775 & 1.578 $\pm$ 0.007 $\pm$ 0.049 & 1.345 $\pm$ 0.002 $\pm$ 0.053 & 1.996 $\pm$ 0.002 $\pm$ 0.087 & 0.3484 $\pm$ 0.0005 $\pm$ 0.0160   \\ \cline{1-5}
 0.825 & 1.606 $\pm$ 0.006 $\pm$ 0.050 & 1.389 $\pm$ 0.002 $\pm$ 0.066 & 2.130 $\pm$ 0.003 $\pm$ 0.117 & 0.3860 $\pm$ 0.0006 $\pm$ 0.0233   \\ \cline{1-5}
 0.875 & 1.625 $\pm$ 0.019 $\pm$ 0.074 & 1.452 $\pm$ 0.005 $\pm$ 0.065 & 2.320 $\pm$ 0.004 $\pm$ 0.122 & 0.4393 $\pm$ 0.0008 $\pm$ 0.0254   \\ \cline{1-5}
 0.925 & 1.649 $\pm$ 0.014 $\pm$ 0.040 & 1.500 $\pm$ 0.005 $\pm$ 0.058 & 2.476 $\pm$ 0.005 $\pm$ 0.119 & 0.4866 $\pm$ 0.0010 $\pm$ 0.0264   \\ \cline{1-5}
 0.975 & 1.669 $\pm$ 0.013 $\pm$ 0.044 & 1.553 $\pm$ 0.005 $\pm$ 0.058 & 2.651 $\pm$ 0.007 $\pm$ 0.113 & 0.5416 $\pm$ 0.0015 $\pm$ 0.0254   \\ \cline{1-5}
 1.025 & 1.673 $\pm$ 0.011 $\pm$ 0.049 & 1.584 $\pm$ 0.004 $\pm$ 0.061 & 2.785 $\pm$ 0.011 $\pm$ 0.116 & 0.5887 $\pm$ 0.0030 $\pm$ 0.0248   \\ \cline{1-5}
 1.075 & 1.706 $\pm$ 0.011 $\pm$ 0.046 & 1.597 $\pm$ 0.004 $\pm$ 0.067 & 2.820 $\pm$ 0.005 $\pm$ 0.140 & 0.6048 $\pm$ 0.0012 $\pm$ 0.0322   \\ \cline{1-5}
 1.125 &                               & 1.648 $\pm$ 0.003 $\pm$ 0.076 & 3.002 $\pm$ 0.005 $\pm$ 0.150 & 0.6627 $\pm$ 0.0013 $\pm$ 0.0370   \\ \cline{1-5}
 1.175 &                               & 1.701 $\pm$ 0.004 $\pm$ 0.055 & 3.179 $\pm$ 0.007 $\pm$ 0.117 & 0.7236 $\pm$ 0.0018 $\pm$ 0.0298   \\ \cline{1-5}
 1.225 & 1.722 $\pm$ 0.009 $\pm$ 0.045 & 1.706 $\pm$ 0.005 $\pm$ 0.066 & 3.245 $\pm$ 0.009 $\pm$ 0.154 & 0.7525 $\pm$ 0.0020 $\pm$ 0.0402   \\ \cline{1-5}
 1.275 & 1.736 $\pm$ 0.006 $\pm$ 0.086 & 1.732 $\pm$ 0.005 $\pm$ 0.060 & 3.364 $\pm$ 0.012 $\pm$ 0.126 & 0.8021 $\pm$ 0.0036 $\pm$ 0.0309   \\ \cline{1-5}
 1.325 & 1.792 $\pm$ 0.015 $\pm$ 0.050 & 1.828 $\pm$ 0.004 $\pm$ 0.076 & 3.561 $\pm$ 0.011 $\pm$ 0.178 & 0.8556 $\pm$ 0.0033 $\pm$ 0.0475   \\ \cline{1-5}
 1.375 & 1.798 $\pm$ 0.027 $\pm$ 0.055 & 1.839 $\pm$ 0.004 $\pm$ 0.082 & 3.630 $\pm$ 0.008 $\pm$ 0.189 & 0.8864 $\pm$ 0.0024 $\pm$ 0.0516   \\ \cline{1-5}
 1.425 & 1.815 $\pm$ 0.007 $\pm$ 0.049 & 1.873 $\pm$ 0.004 $\pm$ 0.073 & 3.741 $\pm$ 0.011 $\pm$ 0.173 & 0.9280 $\pm$ 0.0032 $\pm$ 0.0492   \\ \cline{1-5}
 1.475 & 1.833 $\pm$ 0.006 $\pm$ 0.053 & 1.899 $\pm$ 0.004 $\pm$ 0.073 & 3.839 $\pm$ 0.010 $\pm$ 0.154 & 0.9669 $\pm$ 0.0031 $\pm$ 0.0397   \\ \cline{1-5}
 1.525 & 1.844 $\pm$ 0.008 $\pm$ 0.055 & 1.931 $\pm$ 0.004 $\pm$ 0.082 & 3.968 $\pm$ 0.012 $\pm$ 0.187 & 1.0158 $\pm$ 0.0042 $\pm$ 0.0488   \\ \cline{1-5}
 1.575 & 1.833 $\pm$ 0.006 $\pm$ 0.065 & 1.940 $\pm$ 0.004 $\pm$ 0.096 & 4.022 $\pm$ 0.010 $\pm$ 0.249 & 1.0395 $\pm$ 0.0033 $\pm$ 0.0725   \\ \cline{1-5}
 1.625 & 1.862 $\pm$ 0.020 $\pm$ 0.053 & 1.953 $\pm$ 0.005 $\pm$ 0.091 & 4.116 $\pm$ 0.010 $\pm$ 0.252 & 1.0859 $\pm$ 0.0034 $\pm$ 0.0772   \\ \cline{1-5}
 1.675 &                               & 1.957 $\pm$ 0.005 $\pm$ 0.083 & 4.170 $\pm$ 0.011 $\pm$ 0.231 & 1.1173 $\pm$ 0.0036 $\pm$ 0.0740   \\ \cline{1-5}
 1.725 & 1.857 $\pm$ 0.023 $\pm$ 0.049 & 1.998 $\pm$ 0.005 $\pm$ 0.075 & 4.316 $\pm$ 0.013 $\pm$ 0.218 & 1.1680 $\pm$ 0.0043 $\pm$ 0.0726   \\ \cline{1-5}
 1.775 & 1.884 $\pm$ 0.063 $\pm$ 0.054 & 2.020 $\pm$ 0.011 $\pm$ 0.072 & 4.412 $\pm$ 0.012 $\pm$ 0.194 & 1.2081 $\pm$ 0.0043 $\pm$ 0.0628   \\ \cline{1-5}
 1.825 & 1.862 $\pm$ 0.010 $\pm$ 0.053 & 2.024 $\pm$ 0.006 $\pm$ 0.072 & 4.459 $\pm$ 0.015 $\pm$ 0.168 & 1.2338 $\pm$ 0.0050 $\pm$ 0.0462   \\ \cline{1-5}
 1.875 & 1.837 $\pm$ 0.015 $\pm$ 0.060 & 2.014 $\pm$ 0.006 $\pm$ 0.101 & 4.446 $\pm$ 0.015 $\pm$ 0.256 & 1.2363 $\pm$ 0.0046 $\pm$ 0.0798   \\ \cline{1-5}
 1.925 &                               & 2.026 $\pm$ 0.006 $\pm$ 0.093 & 4.551 $\pm$ 0.015 $\pm$ 0.243 & 1.2903 $\pm$ 0.0047 $\pm$ 0.0755   \\ \cline{1-5}
 1.975 & 1.866 $\pm$ 0.010 $\pm$ 0.059 & 2.027 $\pm$ 0.007 $\pm$ 0.091 & 4.539 $\pm$ 0.018 $\pm$ 0.253 & 1.2857 $\pm$ 0.0058 $\pm$ 0.0824   \\ \cline{1-5}
 2.025 & 1.831 $\pm$ 0.014 $\pm$ 0.046 & 2.037 $\pm$ 0.007 $\pm$ 0.092 & 4.677 $\pm$ 0.020 $\pm$ 0.263 & 1.3480 $\pm$ 0.0069 $\pm$ 0.0867   \\ \cline{1-5}
 2.075 &                               & 2.046 $\pm$ 0.008 $\pm$ 0.084 & 4.699 $\pm$ 0.022 $\pm$ 0.232 & 1.3694 $\pm$ 0.0084 $\pm$ 0.0750   \\ \cline{1-5}
 2.125 & 1.870 $\pm$ 0.022 $\pm$ 0.052 & 2.074 $\pm$ 0.008 $\pm$ 0.092 & 4.825 $\pm$ 0.022 $\pm$ 0.269 & 1.4239 $\pm$ 0.0082 $\pm$ 0.0903   \\ \cline{1-5}
 2.175 & 1.846 $\pm$ 0.013 $\pm$ 0.059 & 2.064 $\pm$ 0.010 $\pm$ 0.098 & 4.850 $\pm$ 0.024 $\pm$ 0.282 & 1.4421 $\pm$ 0.0092 $\pm$ 0.0945   \\ \cline{1-5}
 2.225 &                               & 2.053 $\pm$ 0.012 $\pm$ 0.089 & 4.825 $\pm$ 0.024 $\pm$ 0.267 & 1.4442 $\pm$ 0.0093 $\pm$ 0.0912   \\ \cline{1-5}
 2.275 & 1.852 $\pm$ 0.020 $\pm$ 0.050 & 2.062 $\pm$ 0.008 $\pm$ 0.095 & 4.852 $\pm$ 0.023 $\pm$ 0.271 & 1.4606 $\pm$ 0.0092 $\pm$ 0.0917   \\ \cline{1-5}
 2.325 & 1.859 $\pm$ 0.012 $\pm$ 0.058 & 2.081 $\pm$ 0.009 $\pm$ 0.108 & 4.984 $\pm$ 0.025 $\pm$ 0.291 & 1.5149 $\pm$ 0.0098 $\pm$ 0.0959   \\ \cline{1-5}
 2.375 & 1.867 $\pm$ 0.012 $\pm$ 0.055 & 2.060 $\pm$ 0.008 $\pm$ 0.101 & 4.876 $\pm$ 0.023 $\pm$ 0.275 & 1.4832 $\pm$ 0.0091 $\pm$ 0.0921   \\ \cline{1-5}
 2.425 &                               & 2.051 $\pm$ 0.008 $\pm$ 0.107 & 4.860 $\pm$ 0.023 $\pm$ 0.257 & 1.4835 $\pm$ 0.0089 $\pm$ 0.0850   \\ \cline{1-5}
 2.475 & 1.793 $\pm$ 0.068 $\pm$ 0.089 & 2.056 $\pm$ 0.010 $\pm$ 0.082 & 4.971 $\pm$ 0.020 $\pm$ 0.234 & 1.5362 $\pm$ 0.0079 $\pm$ 0.0796   \\ \cline{1-5}
 2.525 & 1.845 $\pm$ 0.031 $\pm$ 0.066 & 2.035 $\pm$ 0.010 $\pm$ 0.110 & 4.899 $\pm$ 0.018 $\pm$ 0.260 & 1.5176 $\pm$ 0.0063 $\pm$ 0.0751   \\ \cline{1-5}
 2.575 & 1.841 $\pm$ 0.019 $\pm$ 0.052 & 2.050 $\pm$ 0.010 $\pm$ 0.103 & 4.972 $\pm$ 0.021 $\pm$ 0.280 & 1.5556 $\pm$ 0.0078 $\pm$ 0.0915   \\ \cline{1-5}
 2.625 &                               & 2.035 $\pm$ 0.011 $\pm$ 0.122 & 4.884 $\pm$ 0.024 $\pm$ 0.293 & 1.5218 $\pm$ 0.0087 $\pm$ 0.0933   \\ \cline{1-5}
 2.675 &                               & 2.018 $\pm$ 0.009 $\pm$ 0.024 & 4.896 $\pm$ 0.022 $\pm$ 0.277 & 1.5457 $\pm$ 0.0091 $\pm$ 0.0988   \\ \cline{1-5}
 2.725 &                               & 2.028 $\pm$ 0.011 $\pm$ 0.099 & 4.933 $\pm$ 0.025 $\pm$ 0.283 & 1.5634 $\pm$ 0.0094 $\pm$ 0.0970   \\ \cline{1-5}
 2.775 &                               & 2.028 $\pm$ 0.017 $\pm$ 0.107 & 4.931 $\pm$ 0.029 $\pm$ 0.293 & 1.5677 $\pm$ 0.0096 $\pm$ 0.0989   \\ \cline{1-5}
 2.825 & 1.836 $\pm$ 0.026 $\pm$ 0.061 & 2.031 $\pm$ 0.014 $\pm$ 0.118 & 5.004 $\pm$ 0.028 $\pm$ 0.326 & 1.6030 $\pm$ 0.0098 $\pm$ 0.1081   \\ \cline{1-5}
 2.875 & 1.839 $\pm$ 0.016 $\pm$ 0.057 & 2.019 $\pm$ 0.013 $\pm$ 0.108 & 4.976 $\pm$ 0.027 $\pm$ 0.309 & 1.6032 $\pm$ 0.0098 $\pm$ 0.1036   \\ \cline{1-5}
\end{tabular}
\end{ruledtabular}
\end{table*}
\begin{table*}
\begin{ruledtabular}
\begin{tabular}{|c|c|c|c|c|} \cline{1-5}
$Q^2~[$(GeV/c)$^2]$ & $M_2(Q^2)$x$10^{-1}$ & $M_4(Q^2)$x$10^{-2}$ & $M_6(Q^2)$x$10^{-3}$ & $M_8(Q^2)$x$10^{-3}$ \\ \cline{1-5}
 2.925 &                               & 2.023 $\pm$ 0.016 $\pm$ 0.112 & 5.007 $\pm$ 0.033 $\pm$ 0.303 & 1.6219 $\pm$ 0.0104 $\pm$ 0.0970   \\ \cline{1-5}
 2.975 & 1.843 $\pm$ 0.033 $\pm$ 0.050 & 2.018 $\pm$ 0.014 $\pm$ 0.102 & 4.983 $\pm$ 0.027 $\pm$ 0.294 & 1.6145 $\pm$ 0.0092 $\pm$ 0.0941   \\ \cline{1-5}
 3.025 & 1.816 $\pm$ 0.068 $\pm$ 0.058 & 1.978 $\pm$ 0.016 $\pm$ 0.104 & 4.926 $\pm$ 0.027 $\pm$ 0.314 & 1.6042 $\pm$ 0.0090 $\pm$ 0.1057   \\ \cline{1-5}
 3.075 & 1.804 $\pm$ 0.060 $\pm$ 0.055 & 1.992 $\pm$ 0.022 $\pm$ 0.114 & 4.942 $\pm$ 0.040 $\pm$ 0.352 & 1.6136 $\pm$ 0.0119 $\pm$ 0.1222   \\ \cline{1-5}
 3.125 &                               &                               &                               & 1.6293 $\pm$ 0.0118 $\pm$ 0.1491   \\ \cline{1-5}
 3.175 &                               & 2.011 $\pm$ 0.031 $\pm$ 0.141 & 5.002 $\pm$ 0.064 $\pm$ 0.372 & 1.6524 $\pm$ 0.0159 $\pm$ 0.1310   \\ \cline{1-5}
 3.225 &                               & 1.968 $\pm$ 0.021 $\pm$ 0.112 & 4.916 $\pm$ 0.040 $\pm$ 0.358 & 1.6289 $\pm$ 0.0122 $\pm$ 0.1315   \\ \cline{1-5}
 3.275 &                               & 2.007 $\pm$ 0.022 $\pm$ 0.116 & 4.985 $\pm$ 0.043 $\pm$ 0.351 & 1.6478 $\pm$ 0.0127 $\pm$ 0.1304   \\ \cline{1-5}
 3.325 & 1.808 $\pm$ 0.033 $\pm$ 0.080 & 1.979 $\pm$ 0.014 $\pm$ 0.096 & 4.944 $\pm$ 0.032 $\pm$ 0.332 & 1.6321 $\pm$ 0.0120 $\pm$ 0.1264   \\ \cline{1-5}
 3.375 & 1.804 $\pm$ 0.029 $\pm$ 0.055 & 1.981 $\pm$ 0.016 $\pm$ 0.086 & 4.976 $\pm$ 0.031 $\pm$ 0.312 & 1.6543 $\pm$ 0.0119 $\pm$ 0.1243   \\ \cline{1-5}
 3.425 &                               &                               & 5.034 $\pm$ 0.035 $\pm$ 0.325 & 1.6787 $\pm$ 0.0125 $\pm$ 0.1285   \\ \cline{1-5}
 3.475 &                               & 1.943 $\pm$ 0.013 $\pm$ 0.064 & 4.915 $\pm$ 0.032 $\pm$ 0.253 & 1.6489 $\pm$ 0.0118 $\pm$ 0.1079   \\ \cline{1-5}
 3.525 &                               & 1.951 $\pm$ 0.021 $\pm$ 0.088 & 4.999 $\pm$ 0.052 $\pm$ 0.316 & 1.6918 $\pm$ 0.0168 $\pm$ 0.1238   \\ \cline{1-5}
 3.575 &                               &                               & 5.021 $\pm$ 0.043 $\pm$ 0.268 & 1.6858 $\pm$ 0.0142 $\pm$ 0.1049   \\ \cline{1-5}
 3.625 &                               &                               &                               &                                    \\ \cline{1-5}
 3.675 &                               & 1.930 $\pm$ 0.024 $\pm$ 0.040 & 4.857 $\pm$ 0.063 $\pm$ 0.310 & 1.6493 $\pm$ 0.0197 $\pm$ 0.1199   \\ \cline{1-5}
 3.725 &                               &                               &                               & 1.6698 $\pm$ 0.0160 $\pm$ 0.1276   \\ \cline{1-5}
 5.967 &                               & 1.810 $\pm$ 0.015 $\pm$ 0.116 & 4.597 $\pm$ 0.044 $\pm$ 0.553 &                                  \\ \cline{1-5}
 7.268 &                               & 1.743 $\pm$ 0.011 $\pm$ 0.044 &                               &                                  \\ \cline{1-5}
 7.645 &                               &                               & 4.374 $\pm$ 0.044 $\pm$ 0.098 & 1.5659 $\pm$ 0.0202 $\pm$ 0.0396 \\ \cline{1-5}
 8.027 &                               &                               & 4.279 $\pm$ 0.027 $\pm$ 0.135 & 1.5205 $\pm$ 0.0107 $\pm$ 0.0642 \\ \cline{1-5}
 8.434 &                               & 1.653 $\pm$ 0.014 $\pm$ 0.084 & 4.223 $\pm$ 0.032 $\pm$ 0.109 & 1.5264 $\pm$ 0.0122 $\pm$ 0.0419 \\ \cline{1-5}
 8.857 & 1.723 $\pm$ 0.015 $\pm$ 0.041 & 1.645 $\pm$ 0.019 $\pm$ 0.027 & 4.108 $\pm$ 0.042 $\pm$ 0.109 & 1.4712 $\pm$ 0.0138 $\pm$ 0.0566 \\ \cline{1-5}
 9.781 &                               & 1.653 $\pm$ 0.010 $\pm$ 0.061 & 4.130 $\pm$ 0.034 $\pm$ 0.146 & 1.4818 $\pm$ 0.0167 $\pm$ 0.0666 \\ \cline{1-5}
10.267 & 1.752 $\pm$ 0.015 $\pm$ 0.052 & 1.622 $\pm$ 0.019 $\pm$ 0.031 & 4.016 $\pm$ 0.035 $\pm$ 0.095 & 1.4321 $\pm$ 0.0113 $\pm$ 0.0367 \\ \cline{1-5}
10.793 &                               &                               & 3.987 $\pm$ 0.106 $\pm$ 0.761 & 1.4256 $\pm$ 0.0175 $\pm$ 0.1103 \\ \cline{1-5}
11.345 & 1.731 $\pm$ 0.016 $\pm$ 0.041 & 1.573 $\pm$ 0.018 $\pm$ 0.035 & 3.853 $\pm$ 0.041 $\pm$ 0.140 & 1.3644 $\pm$ 0.0176 $\pm$ 0.0793 \\ \cline{1-5}
11.939 & 1.759 $\pm$ 0.008 $\pm$ 0.042 & 1.596 $\pm$ 0.013 $\pm$ 0.031 & 3.910 $\pm$ 0.040 $\pm$ 0.111 & 1.3860 $\pm$ 0.0181 $\pm$ 0.0574 \\ \cline{1-5}
13.185 &                               & 1.525 $\pm$ 0.016 $\pm$ 0.021 & 3.681 $\pm$ 0.029 $\pm$ 0.067 & 1.3011 $\pm$ 0.0091 $\pm$ 0.0336 \\ \cline{1-5}
15.310 & 1.686 $\pm$ 0.014 $\pm$ 0.074 & 1.471 $\pm$ 0.019 $\pm$ 0.032 & 3.533 $\pm$ 0.044 $\pm$ 0.133 &                                  \\ \cline{1-5}
16.902 & 1.718 $\pm$ 0.010 $\pm$ 0.051 & 1.450 $\pm$ 0.017 $\pm$ 0.025 & 3.392 $\pm$ 0.058 $\pm$ 0.073 & 1.1752 $\pm$ 0.0252 $\pm$ 0.0283 \\ \cline{1-5}
18.697 & 1.679 $\pm$ 0.033 $\pm$ 0.097 & 1.401 $\pm$ 0.013 $\pm$ 0.027 & 3.275 $\pm$ 0.039 $\pm$ 0.088 & 1.1377 $\pm$ 0.0147 $\pm$ 0.0346 \\ \cline{1-5}
19.629 &                               &                               &                               &*1.1061 $\pm$ 0.0221 $\pm$ 0.0473 \\ \cline{1-5}
21.625 & 1.677 $\pm$ 0.008 $\pm$ 0.041 &                               &                               &                                  \\ \cline{1-5}
24.192 &*1.711 $\pm$ 0.007 $\pm$ 0.114 & 1.385 $\pm$ 0.008 $\pm$ 0.024 &                               &*1.0751 $\pm$ 0.0143 $\pm$ 0.0433 \\ \cline{1-5}
26.599 &*1.665 $\pm$ 0.062 $\pm$ 0.135 &                               &                               &                                  \\ \cline{1-5}
28.192 &*1.702 $\pm$ 0.009 $\pm$ 0.140 & 1.344 $\pm$ 0.007 $\pm$ 0.037 &                               &*1.0109 $\pm$ 0.0096 $\pm$ 0.0808 \\ \cline{1-5}
31.858 &*1.703 $\pm$ 0.010 $\pm$ 0.096 &                               &                               &                                  \\ \cline{1-5}
36.750 &*1.696 $\pm$ 0.013 $\pm$ 0.111 & 1.314 $\pm$ 0.009 $\pm$ 0.057 & 2.971 $\pm$ 0.027 $\pm$ 0.313 &*1.0027 $\pm$ 0.0135 $\pm$ 0.1906 \\ \cline{1-5}
44.000 &*1.681 $\pm$ 0.013 $\pm$ 0.085 &                               &                               &                                  \\ \cline{1-5}
49.750 &*1.658 $\pm$ 0.019 $\pm$ 0.101 &                               &                               &                                  \\ \cline{1-5}
57.000 &*1.694 $\pm$ 0.017 $\pm$ 0.170 &                               &                               &                                  \\ \cline{1-5}
64.884 &*1.636 $\pm$ 0.043 $\pm$ 0.114 & 1.222 $\pm$ 0.053 $\pm$ 0.044 & 2.708 $\pm$ 0.082 $\pm$ 0.193 &*0.8945 $\pm$ 0.0161 $\pm$ 0.1164 \\ \cline{1-5}
75.000 &                               &*1.206 $\pm$ 0.008 $\pm$ 0.025 &*2.651 $\pm$ 0.024 $\pm$ 0.150 &                                  \\ \cline{1-5}
88.000 &*1.669 $\pm$ 0.088 $\pm$ 0.075 &*1.199 $\pm$ 0.038 $\pm$ 0.035 &*2.630 $\pm$ 0.057 $\pm$ 0.202 &                                  \\ \cline{1-5}
99.000 &                               &*1.179 $\pm$ 0.012 $\pm$ 0.034 &*2.568 $\pm$ 0.029 $\pm$ 0.228 &                                  \\ \cline{1-5}
\end{tabular}
\end{ruledtabular}
\end{table*}

\begin{table*}
\caption{\label{table:twist1}Extracted parameters of the twist
expansion. The contribution of higher twists to $M_2$ was
too small to be extracted by the present procedure.}
\begin{ruledtabular}
\begin{tabular}{|c|c|c|c|c|} \hline
               & $M_2$                            & $M_4$                            & $M_6$                              & $M_8$                             \\ \hline
$A_2$          & 0.174$\pm$ 0.006                 & (1.61$\pm$ 0.04)$\times 10^{-2}$ & (3.98$\pm$ 0.18)$\times 10^{-3}$   & (1.39$\pm$ 0.07)$\times 10^{-3}$  \\ \hline
$a^{(4)}$      & (1.4 $\pm$ 1.8) $\times 10^{-3}$ & (3.6$\pm$ 1.4)$\times 10^{-4}$   & (1.9$\pm$ 0.14)$\times 10^{-4}$    & (1.69$\pm$ 0.16)$\times 10^{-4}$  \\ \hline
$\gamma^{(4)}$ & -                                & 5.7 $\pm$ 0.6                    & 7.4 $\pm$ 0.3                      & 6.2 $\pm$ 0.3                     \\ \hline
$a^{(6)}$      & -                                & (-9.5 $\pm$ 3.4)$\times 10^{-5}$ & (-6.57 $\pm$ 0.53)$\times 10^{-5}$ &(-5.75 $\pm$ 0.44)$\times 10^{-5}$  \\ \hline
$\gamma^{(6)}$ & -                                & 4.4 $\pm$ 0.6                    & 5.7 $\pm$ 0.3                      & 4.6 $\pm$ 0.3                      \\ \hline
\end{tabular}
\end{ruledtabular}
\end{table*}

\begin{table*}
\caption{\label{table:ltw} The extracted leading twist contribution $\eta_n(Q^2)$
(see Eq.~\ref{eq:SGR})
shown in Fig.~\ref{fig:twists}, reported
with systematic errors.}
\begin{ruledtabular}
\begin{tabular}{|c|c|c|c|c|} \cline{1-5}
$Q^2~[$(GeV/c)$^2]$ & $\eta_2(Q^2)$x$10^{-1}$ & $\eta_4(Q^2)$x$10^{-2}$ & $\eta_6(Q^2)$x$10^{-2}$ & $\eta_8(Q^2)$x$10^{-2}$ \\ \cline{1-5}
 1.025 & 2.13 $\pm$ 0.07 & 3.62 $\pm$ 0.09 & 1.665 $\pm$ 0.07 & 1.223 $\pm$ 0.065 \\ \cline{1-5}
 1.075 & 2.11 $\pm$ 0.07 & 3.49 $\pm$ 0.09 & 1.522 $\pm$ 0.07 & 1.022 $\pm$ 0.055 \\ \cline{1-5}
 1.125 & 2.09 $\pm$ 0.07 & 3.38 $\pm$ 0.08 & 1.410 $\pm$ 0.06 & 0.883 $\pm$ 0.047 \\ \cline{1-5}
 1.175 & 2.08 $\pm$ 0.07 & 3.28 $\pm$ 0.08 & 1.319 $\pm$ 0.06 & 0.781 $\pm$ 0.041 \\ \cline{1-5}
 1.225 & 2.07 $\pm$ 0.07 & 3.19 $\pm$ 0.08 & 1.243 $\pm$ 0.05 & 0.704 $\pm$ 0.037 \\ \cline{1-5}
 1.275 & 2.05 $\pm$ 0.07 & 3.11 $\pm$ 0.08 & 1.179 $\pm$ 0.05 & 0.643 $\pm$ 0.034 \\ \cline{1-5}
 1.325 & 2.04 $\pm$ 0.07 & 3.04 $\pm$ 0.07 & 1.125 $\pm$ 0.05 & 0.593 $\pm$ 0.031 \\ \cline{1-5}
 1.375 & 2.03 $\pm$ 0.07 & 2.97 $\pm$ 0.07 & 1.077 $\pm$ 0.05 & 0.553 $\pm$ 0.029 \\ \cline{1-5}
 1.425 & 2.02 $\pm$ 0.07 & 2.91 $\pm$ 0.07 & 1.036 $\pm$ 0.05 & 0.519 $\pm$ 0.027 \\ \cline{1-5}
 1.475 & 2.01 $\pm$ 0.07 & 2.86 $\pm$ 0.07 & 0.999 $\pm$ 0.04 & 0.490 $\pm$ 0.026 \\ \cline{1-5}
 1.525 & 2.00 $\pm$ 0.07 & 2.81 $\pm$ 0.07 & 0.966 $\pm$ 0.04 & 0.465 $\pm$ 0.024 \\ \cline{1-5}
 1.575 & 1.99 $\pm$ 0.07 & 2.76 $\pm$ 0.07 & 0.936 $\pm$ 0.04 & 0.443 $\pm$ 0.023 \\ \cline{1-5}
 1.625 & 1.98 $\pm$ 0.07 & 2.72 $\pm$ 0.07 & 0.910 $\pm$ 0.04 & 0.424 $\pm$ 0.022 \\ \cline{1-5}
 1.675 & 1.97 $\pm$ 0.07 & 2.68 $\pm$ 0.07 & 0.886 $\pm$ 0.04 & 0.407 $\pm$ 0.021 \\ \cline{1-5}
 1.725 & 1.96 $\pm$ 0.07 & 2.64 $\pm$ 0.06 & 0.864 $\pm$ 0.04 & 0.392 $\pm$ 0.021 \\ \cline{1-5}
 1.775 & 1.95 $\pm$ 0.07 & 2.61 $\pm$ 0.06 & 0.844 $\pm$ 0.04 & 0.378 $\pm$ 0.020 \\ \cline{1-5}
 1.825 & 1.95 $\pm$ 0.07 & 2.57 $\pm$ 0.06 & 0.825 $\pm$ 0.04 & 0.366 $\pm$ 0.019 \\ \cline{1-5}
 1.875 & 1.94 $\pm$ 0.07 & 2.54 $\pm$ 0.06 & 0.808 $\pm$ 0.04 & 0.355 $\pm$ 0.019 \\ \cline{1-5}
 1.925 & 1.93 $\pm$ 0.07 & 2.51 $\pm$ 0.06 & 0.792 $\pm$ 0.03 & 0.344 $\pm$ 0.018 \\ \cline{1-5}
 1.975 & 1.93 $\pm$ 0.07 & 2.49 $\pm$ 0.06 & 0.777 $\pm$ 0.03 & 0.335 $\pm$ 0.018 \\ \cline{1-5}
 2.025 & 1.92 $\pm$ 0.07 & 2.46 $\pm$ 0.06 & 0.763 $\pm$ 0.03 & 0.326 $\pm$ 0.017 \\ \cline{1-5}
 2.075 & 1.91 $\pm$ 0.07 & 2.44 $\pm$ 0.06 & 0.750 $\pm$ 0.03 & 0.318 $\pm$ 0.017 \\ \cline{1-5}
 2.125 & 1.91 $\pm$ 0.07 & 2.41 $\pm$ 0.06 & 0.738 $\pm$ 0.03 & 0.311 $\pm$ 0.016 \\ \cline{1-5}
 2.175 & 1.90 $\pm$ 0.07 & 2.39 $\pm$ 0.06 & 0.726 $\pm$ 0.03 & 0.304 $\pm$ 0.016 \\ \cline{1-5}
 2.225 & 1.90 $\pm$ 0.07 & 2.37 $\pm$ 0.06 & 0.715 $\pm$ 0.03 & 0.298 $\pm$ 0.016 \\ \cline{1-5}
 2.275 & 1.89 $\pm$ 0.07 & 2.35 $\pm$ 0.06 & 0.706 $\pm$ 0.03 & 0.292 $\pm$ 0.015 \\ \cline{1-5}
 2.325 & 1.89 $\pm$ 0.07 & 2.33 $\pm$ 0.06 & 0.697 $\pm$ 0.03 & 0.287 $\pm$ 0.015 \\ \cline{1-5}
 2.375 & 1.89 $\pm$ 0.07 & 2.32 $\pm$ 0.06 & 0.689 $\pm$ 0.03 & 0.283 $\pm$ 0.015 \\ \cline{1-5}
 2.425 & 1.88 $\pm$ 0.07 & 2.30 $\pm$ 0.06 & 0.682 $\pm$ 0.03 & 0.279 $\pm$ 0.014 \\ \cline{1-5}
 2.475 & 1.88 $\pm$ 0.07 & 2.28 $\pm$ 0.06 & 0.675 $\pm$ 0.03 & 0.275 $\pm$ 0.014 \\ \cline{1-5}
 2.525 & 1.88 $\pm$ 0.06 & 2.27 $\pm$ 0.06 & 0.668 $\pm$ 0.03 & 0.271 $\pm$ 0.014 \\ \cline{1-5}
 2.575 & 1.88 $\pm$ 0.06 & 2.25 $\pm$ 0.06 & 0.661 $\pm$ 0.03 & 0.267 $\pm$ 0.014 \\ \cline{1-5}
 2.625 & 1.87 $\pm$ 0.06 & 2.24 $\pm$ 0.05 & 0.654 $\pm$ 0.03 & 0.264 $\pm$ 0.014 \\ \cline{1-5}
 2.675 & 1.87 $\pm$ 0.06 & 2.23 $\pm$ 0.05 & 0.648 $\pm$ 0.03 & 0.260 $\pm$ 0.013 \\ \cline{1-5}
 2.725 & 1.87 $\pm$ 0.06 & 2.21 $\pm$ 0.05 & 0.642 $\pm$ 0.03 & 0.257 $\pm$ 0.013 \\ \cline{1-5}
 2.775 & 1.87 $\pm$ 0.06 & 2.20 $\pm$ 0.05 & 0.637 $\pm$ 0.03 & 0.254 $\pm$ 0.013 \\ \cline{1-5}
 2.825 & 1.86 $\pm$ 0.06 & 2.19 $\pm$ 0.05 & 0.631 $\pm$ 0.03 & 0.251 $\pm$ 0.013 \\ \cline{1-5}
 2.875 & 1.86 $\pm$ 0.06 & 2.18 $\pm$ 0.05 & 0.626 $\pm$ 0.03 & 0.248 $\pm$ 0.013 \\ \cline{1-5}
 2.925 & 1.86 $\pm$ 0.06 & 2.17 $\pm$ 0.05 & 0.621 $\pm$ 0.03 & 0.245 $\pm$ 0.013 \\ \cline{1-5}
 2.975 & 1.86 $\pm$ 0.06 & 2.15 $\pm$ 0.05 & 0.616 $\pm$ 0.03 & 0.243 $\pm$ 0.013 \\ \cline{1-5}
 3.025 & 1.85 $\pm$ 0.06 & 2.14 $\pm$ 0.05 & 0.611 $\pm$ 0.03 & 0.240 $\pm$ 0.012 \\ \cline{1-5}
 3.075 & 1.85 $\pm$ 0.06 & 2.13 $\pm$ 0.05 & 0.606 $\pm$ 0.03 & 0.238 $\pm$ 0.012 \\ \cline{1-5}
 3.125 & 1.85 $\pm$ 0.06 & 2.12 $\pm$ 0.05 & 0.602 $\pm$ 0.03 & 0.236 $\pm$ 0.012 \\ \cline{1-5}
 3.175 & 1.85 $\pm$ 0.06 & 2.11 $\pm$ 0.05 & 0.598 $\pm$ 0.03 & 0.233 $\pm$ 0.012 \\ \cline{1-5}
 3.225 & 1.85 $\pm$ 0.06 & 2.10 $\pm$ 0.05 & 0.593 $\pm$ 0.03 & 0.231 $\pm$ 0.012 \\ \cline{1-5}
 3.275 & 1.84 $\pm$ 0.06 & 2.09 $\pm$ 0.05 & 0.589 $\pm$ 0.03 & 0.229 $\pm$ 0.012 \\ \cline{1-5}
 3.325 & 1.84 $\pm$ 0.06 & 2.08 $\pm$ 0.05 & 0.585 $\pm$ 0.03 & 0.227 $\pm$ 0.012 \\ \cline{1-5}
 3.375 & 1.84 $\pm$ 0.06 & 2.08 $\pm$ 0.05 & 0.582 $\pm$ 0.03 & 0.225 $\pm$ 0.012 \\ \cline{1-5}
 3.425 & 1.84 $\pm$ 0.06 & 2.07 $\pm$ 0.05 & 0.578 $\pm$ 0.03 & 0.223 $\pm$ 0.011 \\ \cline{1-5}
 3.475 & 1.84 $\pm$ 0.06 & 2.06 $\pm$ 0.05 & 0.574 $\pm$ 0.03 & 0.221 $\pm$ 0.011 \\ \cline{1-5}
 3.525 & 1.84 $\pm$ 0.06 & 2.05 $\pm$ 0.05 & 0.571 $\pm$ 0.03 & 0.220 $\pm$ 0.011 \\ \cline{1-5}
 3.575 & 1.83 $\pm$ 0.06 & 2.04 $\pm$ 0.05 & 0.567 $\pm$ 0.03 & 0.218 $\pm$ 0.011 \\ \cline{1-5}
 3.625 & 1.83 $\pm$ 0.06 & 2.03 $\pm$ 0.05 & 0.564 $\pm$ 0.02 & 0.216 $\pm$ 0.011 \\ \cline{1-5}
 3.675 & 1.83 $\pm$ 0.06 & 2.03 $\pm$ 0.05 & 0.561 $\pm$ 0.02 & 0.215 $\pm$ 0.011 \\ \cline{1-5}
 3.725 & 1.83 $\pm$ 0.06 & 2.02 $\pm$ 0.05 & 0.558 $\pm$ 0.02 & 0.213 $\pm$ 0.011 \\ \cline{1-5}
 5.967 & 1.78 $\pm$ 0.06 & 1.79 $\pm$ 0.04 & 0.467 $\pm$ 0.02 & 0.169 $\pm$ 0.009 \\ \cline{1-5}
 7.268 & 1.77 $\pm$ 0.06 & 1.72 $\pm$ 0.04 & 0.438 $\pm$ 0.02 & 0.156 $\pm$ 0.008 \\ \cline{1-5}
 7.645 & 1.76 $\pm$ 0.06 & 1.70 $\pm$ 0.04 & 0.431 $\pm$ 0.02 & 0.153 $\pm$ 0.008 \\ \cline{1-5}
 8.027 & 1.76 $\pm$ 0.06 & 1.68 $\pm$ 0.04 & 0.424 $\pm$ 0.02 & 0.150 $\pm$ 0.008 \\ \cline{1-5}
\end{tabular}
\end{ruledtabular}
\end{table*}
\begin{table*}
\begin{ruledtabular}
\begin{tabular}{|c|c|c|c|c|} \cline{1-5}
$Q^2~[$(GeV/c)$^2]$ & $\eta_2(Q^2)$ & $\eta_4(Q^2)$x$10^{-2}$ & $\eta_6(Q^2)$x$10^{-2}$ & $\eta_8(Q^2)$x$10^{-2}$ \\ \cline{1-5}
 8.434 & 1.75 $\pm$ 0.06 & 1.66 $\pm$ 0.04 & 0.418 $\pm$ 0.02 & 0.147 $\pm$ 0.008 \\ \cline{1-5}
 8.857 & 1.75 $\pm$ 0.06 & 1.65 $\pm$ 0.04 & 0.412 $\pm$ 0.02 & 0.144 $\pm$ 0.008 \\ \cline{1-5}
 9.781 & 1.74 $\pm$ 0.06 & 1.61 $\pm$ 0.04 & 0.400 $\pm$ 0.02 & 0.139 $\pm$ 0.008 \\ \cline{1-5}
10.267 & 1.74 $\pm$ 0.06 & 1.60 $\pm$ 0.04 & 0.395 $\pm$ 0.02 & 0.137 $\pm$ 0.007 \\ \cline{1-5}
10.793 & 1.74 $\pm$ 0.06 & 1.58 $\pm$ 0.04 & 0.389 $\pm$ 0.02 & 0.134 $\pm$ 0.007 \\ \cline{1-5}
11.345 & 1.73 $\pm$ 0.06 & 1.57 $\pm$ 0.04 & 0.384 $\pm$ 0.02 & 0.132 $\pm$ 0.007 \\ \cline{1-5}
11.939 & 1.73 $\pm$ 0.06 & 1.55 $\pm$ 0.04 & 0.379 $\pm$ 0.02 & 0.130 $\pm$ 0.007 \\ \cline{1-5}
13.185 & 1.72 $\pm$ 0.06 & 1.52 $\pm$ 0.04 & 0.369 $\pm$ 0.02 & 0.126 $\pm$ 0.007 \\ \cline{1-5}
15.310 & 1.71 $\pm$ 0.06 & 1.48 $\pm$ 0.04 & 0.355 $\pm$ 0.02 & 0.120 $\pm$ 0.007 \\ \cline{1-5}
16.902 & 1.71 $\pm$ 0.06 & 1.46 $\pm$ 0.04 & 0.347 $\pm$ 0.02 & 0.116 $\pm$ 0.006 \\ \cline{1-5}
18.697 & 1.70 $\pm$ 0.06 & 1.43 $\pm$ 0.04 & 0.338 $\pm$ 0.01 & 0.113 $\pm$ 0.006 \\ \cline{1-5}
19.629 & 1.70 $\pm$ 0.06 & 1.42 $\pm$ 0.03 & 0.334 $\pm$ 0.01 & 0.111 $\pm$ 0.006 \\ \cline{1-5}
21.625 & 1.69 $\pm$ 0.06 & 1.40 $\pm$ 0.03 & 0.327 $\pm$ 0.01 & 0.108 $\pm$ 0.006 \\ \cline{1-5}
24.192 & 1.69 $\pm$ 0.06 & 1.37 $\pm$ 0.03 & 0.319 $\pm$ 0.01 & 0.105 $\pm$ 0.006 \\ \cline{1-5}
26.599 & 1.68 $\pm$ 0.06 & 1.35 $\pm$ 0.03 & 0.312 $\pm$ 0.01 & 0.102 $\pm$ 0.006 \\ \cline{1-5}
28.192 & 1.68 $\pm$ 0.06 & 1.34 $\pm$ 0.03 & 0.309 $\pm$ 0.01 & 0.101 $\pm$ 0.006 \\ \cline{1-5}
31.858 & 1.68 $\pm$ 0.06 & 1.32 $\pm$ 0.03 & 0.301 $\pm$ 0.01 & 0.098 $\pm$ 0.005 \\ \cline{1-5}
36.750 & 1.68 $\pm$ 0.06 & 1.29 $\pm$ 0.03 & 0.293 $\pm$ 0.01 & 0.095 $\pm$ 0.005 \\ \cline{1-5}
44.000 & 1.67 $\pm$ 0.06 & 1.26 $\pm$ 0.03 & 0.283 $\pm$ 0.01 & 0.091 $\pm$ 0.005 \\ \cline{1-5}
49.750 & 1.67 $\pm$ 0.06 & 1.24 $\pm$ 0.03 & 0.277 $\pm$ 0.01 & 0.089 $\pm$ 0.005 \\ \cline{1-5}
57.000 & 1.67 $\pm$ 0.06 & 1.22 $\pm$ 0.03 & 0.270 $\pm$ 0.01 & 0.086 $\pm$ 0.005 \\ \cline{1-5}
64.884 & 1.67 $\pm$ 0.06 & 1.20 $\pm$ 0.03 & 0.264 $\pm$ 0.01 & 0.084 $\pm$ 0.004 \\ \cline{1-5}
75.000 & 1.66 $\pm$ 0.06 & 1.17 $\pm$ 0.03 & 0.257 $\pm$ 0.01 & 0.081 $\pm$ 0.004 \\ \cline{1-5}
88.000 & 1.66 $\pm$ 0.06 & 1.15 $\pm$ 0.03 & 0.250 $\pm$ 0.01 & 0.078 $\pm$ 0.004 \\ \cline{1-5}
99.000 & 1.66 $\pm$ 0.06 & 1.13 $\pm$ 0.03 & 0.245 $\pm$ 0.01 & 0.077 $\pm$ 0.004 \\ \cline{1-5}
\end{tabular}
\end{ruledtabular}
\end{table*}

\begin{figure*}
\includegraphics{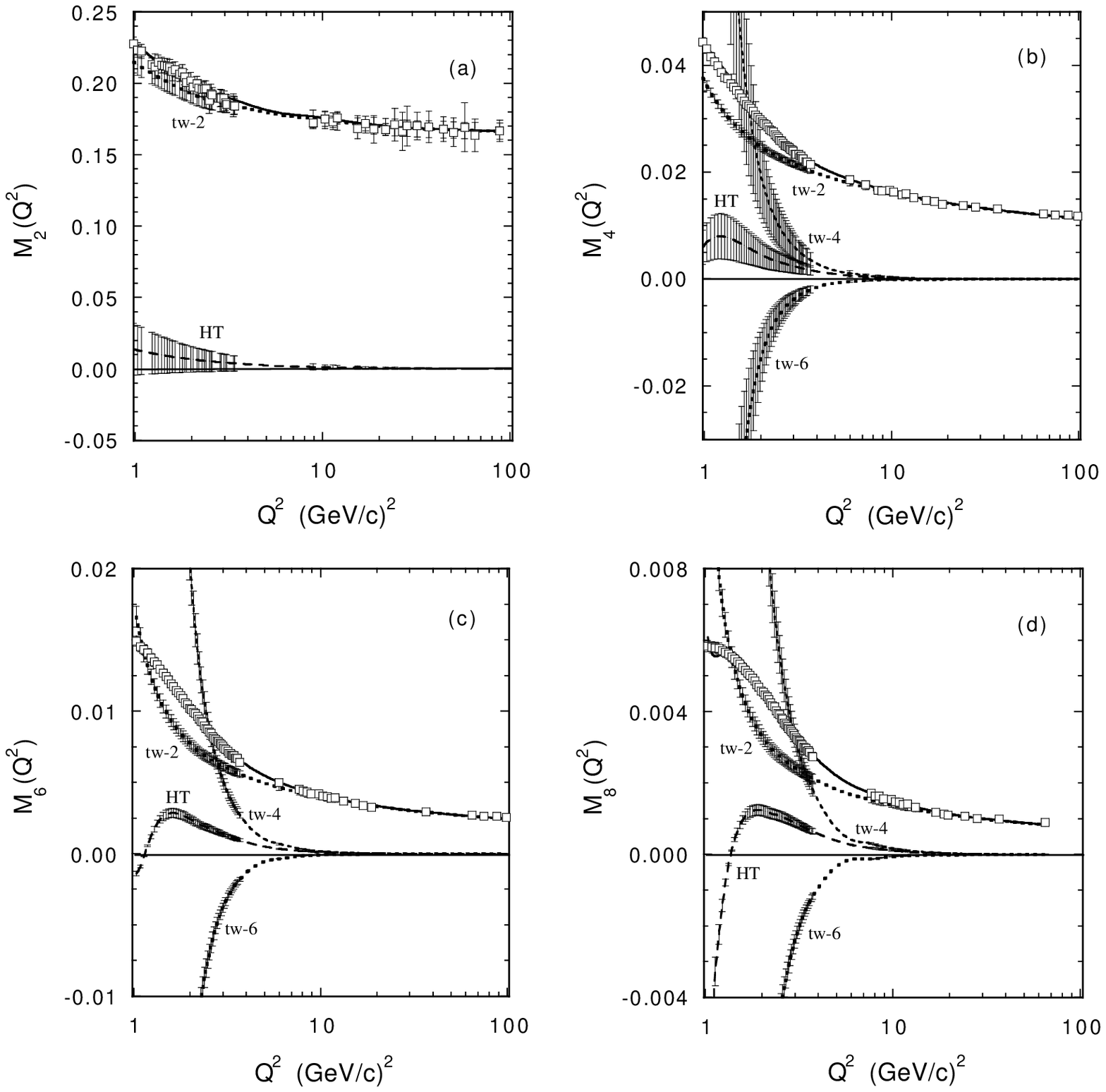}
\caption{\label{fig:twists}
Results of the twist analysis.  The open squares
represent the Nachtman moments obtained in this analysis.  The
solid line is the fit to the moments using Eq.~\ref{eq:twists} with the
parameters listed in Table~\ref{table:twist1}.  The twist-2, twist-4, twist-6
and higher twist (HT) contributions to the fit are indicated.
The twist-2 contribution was calculated using Eq.~\ref{eq:SGR}.}
\end{figure*}

\begin{figure}
\includegraphics[scale=0.8]{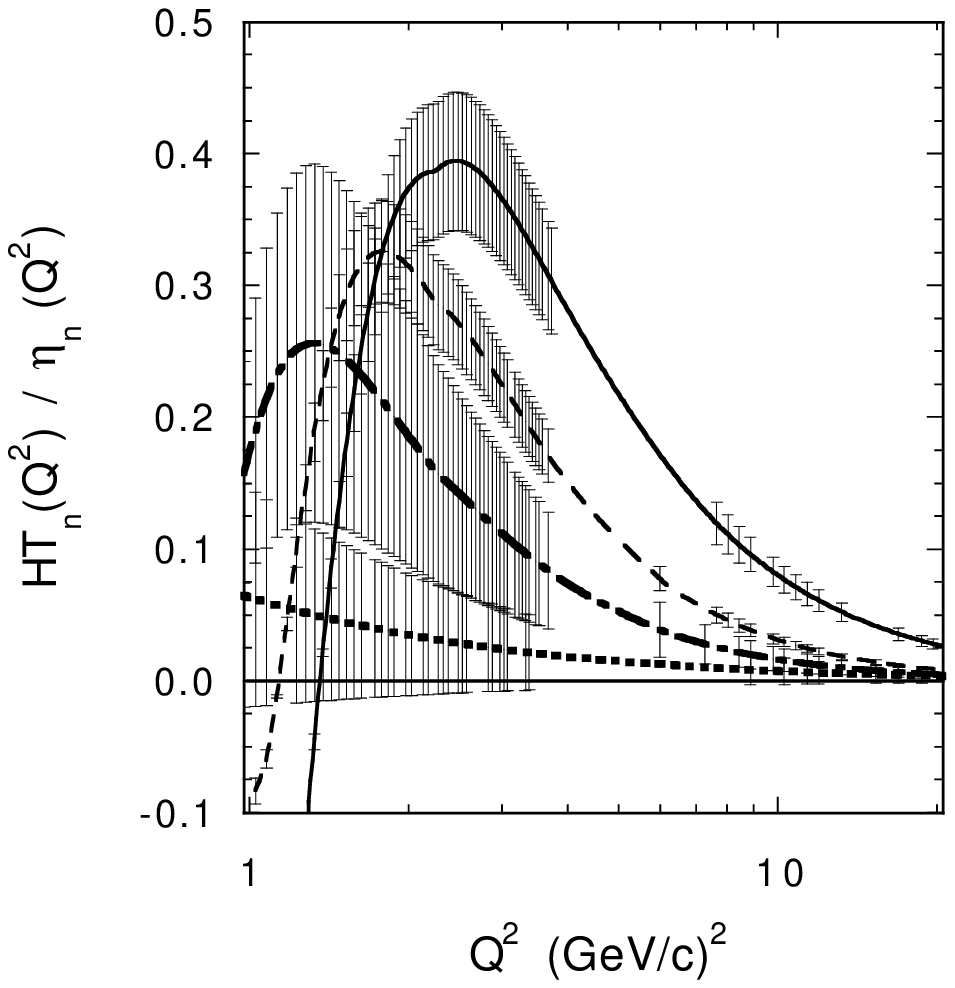}
\caption{\label{fig:ratio}Ratio of the total higher-twist (see Eq.~\ref{eq:HT}) to
the leading twist given in Eq.~\ref{eq:SGR}. Dotted line - $M_2$; Dash-dotted line - $M_4$;
dashed line - $M_6$; continuous line - $M_8$.}
\end{figure}

Our results, including the uncertainties for each twist term
separately, are reported in Fig.~\ref{fig:twists} for $n \geq 2$,
while the ratio of the total higher-twist contribution to the
leading twist is shown in Fig.~\ref{fig:ratio}.
In addition, the extracted leading twist contribution is
reported in Table~\ref{table:ltw}.
It can be seen that:
\begin{itemize}
\item the extracted twist-2 term yields an important contribution
in the whole $Q^2$-range of the present analysis;
\item the $Q^2$-behaviour of the data leaves room for a higher-twist
contribution positive at large $Q^2$ and negative at
$Q^2 \sim 1 - 2$~(GeV/c)$^2$; such a change of sign requires
in Eq.~\ref{eq:HT} a twist-6 term with a sign opposite to
that of the twist-4. As already noted in Refs.~\cite{Ricco1,SIM00},
such opposite signs make
the
total higher-twist contribution
smaller than its individual terms;
\item the total higher-twist contribution is significant at
$Q^2 \approx$ few (GeV/c)$^2$, but it is less than
$\simeq 20$\% of the leading twist for $Q^2 > 5$~(GeV/c)$^2$. 
\end{itemize}

\section{Conclusions}\label{sec:Conclusions}
We extracted the $F_2$ structure function in a continuous
two-dimensional range of $Q^2$ and $x$ from the inclusive
cross section measured with the CLAS detector. Using these
data, together with the previously available world data set,
we evaluated the Nachtmann moments $M_2(Q^2,x)$, $M_4(Q^2,x)$,
$M_6(Q^2,x)$ and $M_8(Q^2,x)$ in the $Q^2$ range
$0.05 - 100$~(GeV/c)$^2$. The present data set covers a
large interval in $x$ thus reducing the uncertainties in the
integration procedure.
The Nachtmann moments obtained in this work have been analyzed
in terms of a twist expansion in order to simultaneously extract
both the leading and the higher twists. The former has not been
treated at fixed order in perturbation theory, but
higher-order corrections of pQCD
were taken into account by means of soft
gluon resummation techniques. Higher twists have been treated
phenomenologically by introducing {\em effective} anomalous
dimensions. The range of the analysis was quite large, ranging
from $1$ to $100$~(GeV/c)$^2$. The leading twist is determined
with a few percent uncertainty, while the precision of the
higher twists increases with $n$ reaching an overall $10$\%
for $n=6$ and $8$,
thanks to the remarkable quality of the experimental moments.

The main results of our twist analysis can be summarized as
follows:
~ i) the contribution of the leading twist calculated in the frame
of pQCD at NLO remains dominant down to $2Q^2/n \sim 1$ (GeV/c)$^2$,
where $n$ is the moment order.
This leads to the conclusion that a pQCD-based description of
the proton structure is relevant also at low $Q^2$, with
significant but not crucial corrections;
~ ii) the total contribution of the multi-parton correlation
effects is not negligible for $Q^2 < 5$~(GeV/c)$^2$ and large $x$
corresponding to the resonance region.
This can be seen by comparing the higher twist contribution to
$M_8$, which is more heavily weighted in $x$, to $M_2$;
~ iii) different higher twist terms tend to compensate each other
in such a way that their sum is small even in a $Q^2$ region
where their absolute contributions exceed the leading twist.
This cancellation is responsible for the duality phenomena and leads
to the prevailing DIS-inspired picture of
photon-proton collisions at low $Q^2$.

Therefore, we demonstrated that a precise determination
of higher twists is feasible with the high quality of the new
CLAS data. The main limitation of the present analysis
is the use of a phenomenological ansatz for the higher twists.
In this respect it is necessary to have better
theoretical knowledge of the renormalization group behaviour
of the relevant higher-twist operators. This would
directly test QCD in its non-perturbative regime through the
comparison of predictions obtained from lattice simulations with these data.

\begin{acknowledgments}
This work was supported by the Istituto Nazionale di Fisica Nucleare,
the French Commissariat \`a l'Energie Atomique, 
French Centre National de la Recherche Scientifique,
the U.S. Department of Energy and National Science Foundation and
the Korea Science and Engineering Foundation.
The Southeastern Universities Research
Association (SURA) operates the Thomas Jefferson National Accelerator
Facility for the United States
Department of Energy under contract DE-AC05-84ER40150.
  
\end{acknowledgments}

\appendix

\section{Fit of the ratio $R\equiv \sigma_L / \sigma_T$}\label{appendix1}
The function $R(x,Q^2)= \sigma_L/\sigma_T$ was described as:
\begin{eqnarray}\nonumber
&&R(x,Q^2)= \\
&&\left\{
\begin{array}{ll}
\frac{(1-x)^3}{(1-x_{th})^3}\bigl[
\frac{0.041\xi_{th}}{\zeta}+
\frac{0.592}{Q^2}-\frac{0.331}{(0.09+Q^4)}\bigr] & W < 2.5 ~, \\ \\
\frac{0.041\xi}{\zeta}+
\frac{0.592}{Q^2}-\frac{0.331}{(0.09+Q^4)} & W > 2.5 ~
\end{array}
\right.
\label{eq:d_sf3}
\end{eqnarray}

This parametrization of $R(x,Q^2)$ consists of two different parts:
the fit for the DIS region ($W > 2.5$ GeV)~\cite{Whitlow_R,bar} and
the function, adjusted to scarce data at small $Q^2$
\cite{BurkertPP,Ratio1,Ratio2}, in the resonance region ($W < 2.5$ GeV).
The systematic error on this parametrization was estimated according to Ref.~\cite{Ricco1}
as follows:
\begin{equation}
\delta_R= \left\{
\begin{array}{ll}
0.08 & W < 2.5 ~,\\ \\
\frac{0.006\xi}{\zeta}+
\frac{0.01}{Q^2}-\frac{0.01}{(0.09+Q^4)} & W > 2.5 ~
\end{array}
\right.
\label{eq:d_sf4}
\end{equation}
\noindent
where
\begin{eqnarray}
&& \zeta = \log{\frac{Q^2}{0.04}} \\ \nonumber
&& \xi = 1+12\frac{Q^2}{1+Q^2}
\frac{0.015625}{0.015625+x^2} \\ \nonumber
&& \xi_{th} = \xi(W=2.5) \\ \nonumber
&& x_{th} = x(W=2.5) \nonumber
\label{eq:d_sf5}
\end{eqnarray}

All dimensional variables are given in GeV.

\end{document}